\documentclass[aip, jcp,
 amsmath,amssymb,
 reprint,%
]{revtex4-1}
\usepackage{changes}
\usepackage{dcolumn}% Align table columns on decimal point
\usepackage{bm}% bold math
\usepackage[utf8]{inputenc}
\usepackage[T1]{fontenc}
\usepackage{mathptmx}
\usepackage{etoolbox}
\usepackage{graphicx} %package for adding images.
\graphicspath{}
\usepackage{tabularx} %for table
\usepackage{paralist}%in para list
\usepackage{amsmath} %math
\makeatletter
\def\@email#1#2{%
 \endgroup
 \patchcmd{\titleblock@produce}
  {\frontmatter@RRAPformat}
  {\frontmatter@RRAPformat{\produce@RRAP{*#1\href{mailto:#2}{#2}}}\frontmatter@RRAPformat}
  {}{}
}%
\makeatother
\begin{document}
\preprint{AIP/123-QED} 
\title{\bfseries Mixing small proteins with
  lipids and cholesterol}

\author{Subhadip Basu}

\affiliation{Department of Biomedical Engineering,\@ Ben Gurion
  University of the Negev\\ Be\rq er Sheva 84105, Israel}

\author{Oded Farago}

\affiliation{Department of Biomedical Engineering,\@ Ben Gurion
  University of the Negev\\ Be\rq er Sheva 84105, Israel}
\email{ofarago@bgu.ac.il}
\date{\today}

\begin{abstract}

Many ternary mixtures composed of saturated and unsaturated lipids
with cholesterol (Chol) exhibit a region of
coexistence between liquid-disordered ($L_d$) and liquid-ordered
($L_o$) domains, bearing some similarities to lipid 
rafts in biological membranes. However, biological rafts also contain many proteins that
interact with the lipids and modify the distirubtion of lipids. 
Here, we extend a previously published lattice model of
ternary DPPC/DOPC/Chol mixtures by introducing a small amount of small
proteins (peptides). We use Monte Carlo simulations to explore 
the mixing phase behavior of the components as a function of the
interaction parameter representing the affinity between the proteins
and the saturated DPPC chains, and for different mixture
compositions. At moderate fractions of DPPC, the system is in a
two-phase $L_d+L_o$ coexistence, and the proteins exhibit a simple
partition behavior between the phases that depends on the
protein-lipid affinity parameter. At low DPPC compositions, the
mixture is in $L_d$ phase with local nanoscopic ordered
domains. Addition of proteins with sufficiently strong attraction to
the saturated lipids can induce the separation of a distinct $L_o$
large domain with tightly-packed gel-like clusters of proteins and
saturated lipids. Consistent with the theory of phase transitions, we observe that   
the domain sizes grow when the mixture composition is in the vicinity of the critical point.  
Our simulations show that the addition of a small amount of proteins to such mixtures can 
cause their size to grow even further, and  lead to the formation of metastable dynamic $L_o$ 
domains with sizes comparable to biological rafts.
\end{abstract}
\maketitle
\section{Introduction}

Cell membranes are thin bilayer sheets that define the boundaries of
cells and their internal organelles. In addition to protecting the
cells, membranes also play important roles in various cellular
functions like signal transduction, molecular transports, and
molecular organization of cellular
processes.\cite{Alberts2017,cheng2019biological} Composition wise,
biological membranes consist of hundreds of different types of
lipids\cite{shevchenko2010lipidomics,SHEN2014359}, which can be
divided into three main categories: phospholipids, glycolipids and
sterols.\cite{Alberts2017,cheng2019biological,BLANCO2017215} While the
lipids constitute the main element of the cell membrane, about~50\% of
the area of the membrane is occupied by
proteins.\cite{Alberts2017,guigas2016effects}
%and three classes of proteins can be found in a membrane:integral, peripheral and lipid-anchored \cite{alberts2014molecular}.
Many of the membrane functions like cell-to cell communication and
active transport of molecules are associated with these
proteins.\cite{konings1996transport,jelokhani2022membrane,whitford2013proteins}

Lateral organization of lipids in the membrane leaflets and its
correlation to the membrane functionalities has been a topic of
scientific interest for several decades. Back in 1972, the
fluid-mosaic model was proposed, describing the cell membranes as a
random mixture of lipids with proteins embedded
within.\cite{sj1972fluid} Soon later, it was realized experimentally
\cite{ahmed1997origin,brown1992sorting,pralle2000sphingolipid} that
although entropy favors random mixing, interactions between different
types of lipids may promote the formation of membrane domains with
different lipid
compositions.\cite{cheng2019biological,ackerman2015lipid,huang1999microscopic}
This has led to the development of the lipid raft
hypothesis:\cite{schuck2003resistance} Rafts are defined as
heterogeneous, dynamic, cholesterol and sphingolipid-enriched membrane
domains (10-200 nm), with a potential to form microscopic domains
($>300\,\text{nm}$) in presence of proteins.\cite{pike2006rafts} Rafts
are liquid-ordered ($L_o$) domains, having properties intermediate
between the liquid-disordered ($L_d$) and gel ($S_o$)
phases.\cite{mouritsen2010liquid} Similarly to the former, the lipids
in the liquid-ordered domains are mobile and free to diffuse in the
membrane plane.\cite{filippov2004lipid} However, their hydrocarbon
chains are ordered, fully extended and tightly packed, as in the gel
phase.\cite{holl2008cell}
%The exact composition, nature of lipid rafts are still debated.

One of the potential routes to investigate the lateral arrangement of
lipids in the cell membrane is to map out its phase
diagram. Unfortunately, the structural complexity of real biological
membranes makes such a task almost impossible. Therefore, efforts have
been made to establish the phase diagram of compositionally simpler
model systems, especially of ternary lipid mixtures composed of a
unsaturated low melting temperature lipid (like DOPC), a saturated
high melting temperature lipid (like DPPC) and cholesterol
(Chol).\cite{marsh2013handbook} Investigations of many such ternary
mixtures revealed regions of phase coexistence between the $L_o$
phase, dominated by saturated lipids and Chol population, and a $L_d$
phase, composed mainly of unsaturated
lipids.\cite{feigenson2009phase,komura2014physical,veatch2007critical,hirst2011phase}
Depending on the identity of the lipids and temperature, the
liquid-liquid phase separation may be macroscopic-thermodynamic (Type
II mixtures) or local (Type I mixtures).\cite{feigenson2009phase} The
latter case seems to be more relevant to lipid rafts in complex
biological membranes which, as noted above, are of typical size of
several tens of nanometers. At high fractions of the saturated lipids,
the phase diagrams of ternary mixtures also include regions of
coexistence between the liquid phases and the gel $S_o$ phase, where
the saturated lipids are immobile and very tightly packed.

Several experimental techniques like fluorescence
microscopy,\cite{stockl2010detection,bagatolli2001direct,bagatolli2000correlation}
F\"orster Resonance Energy Transfer
(FRET),\cite{farago2021beginner,koukalova2017lipid,de2005lipid,vsachl2011limitations}
Interferometric Scattering
Microscopy,\cite{de2015dynamic,giocondi2001phase} Atomic Force
Microscopy
(AFM),\cite{tokumasu2003nanoscopic,choucair2007preferential} Nuclear
Magnetic Resonance (NMR)\cite{veatch2004liquid,veatch2007critical}
have been successfully applied to detect lipid domains.  Apart from
experimental means, Molecular Dynamics (MD) and Monte Carlo (MC)
simulations have been extensively used to investigate lipid domain
formation.\cite{marrink2019computational,BENNETT20131765,bennett2013computer}
All-atom MD simulations revealed many structural details of lipid
domains like the presence of sub-structures within
them.\cite{javanainen2017nanoscale,sodt2014molecular,sodt2015hexagonal}
The role of sterols in packing of lipids have also been investigated
using all-atom MD simulation.\cite{cournia2007differential} However,
because of the large temporal and spatial scales associated with the
process, most atomistic simulations of lipid mixtures capture only the
onset of the formation of liquid ordered
domains.\cite{hakobyan2013phase,bennett2018phospholipid,gu2020phase,BENNETT20131765}
To access larger length- and time-scales, coarse-grained (CG)
simulations have been employed to observe phase separation,
\cite{perlmutter2009inhibiting,rosetti2012comparison,davis2013predictions,baoukina2013computer,pantelopulos2018regimes}
and the phase diagrams of different ternary lipid mixtures have been
determined and found to be in good agreement with experimental
findings.\cite{arnarez2016hysteresis,wang2016dppc,he2018identifying,carpenter2018capturing,podewitz2018phase}
Besides atomistic and CG simulations of specific mixtures, ultra CG
and lattice models have been also developed to probe phase separation
phenomena in lipid
mixtures.\cite{PhysRevResearch.3.L042030,D2SM01025A,meinhardt2019structure,almeida2011simple}
Such models facilitate simulations of complete phase separation in
very large systems, and their simplicity can help characterizing the
mechanisms governing the thermodynamic behavior of lipid-Chol
mixtures.

Proteins may show affinity to specific lipid domains which has
implications in many cellular phenomena. Certain
glycosylphosphatidylinositol (GPI)-anchored proteins are associated
with lipid rafts.\cite{lingwood2010lipid} B-cell receptor (BCR)
proteins tend to aggregate in the $L_o$ domains, and such preferential
localization was reported to facilitate BCR
activation.\cite{stone2017protein} Segregation property of another
similar protein, namely T-cell receptors (TCR), remain controversial,
as its affinity toward both $L_o$ and $L_d$ domains has been
reported.\cite{danylchuk2020redesigning,beck2015nanoclusters} Protein
partitioning in the neuronal membrane is also of significant
importance. There are reports on the raft-dependent functionality of
neurotransmitters like choline and
serotonin.\cite{magnani2004partitioning,cuddy2014regulation} Some
nerve growth factors, like trkA and p75, also prefer to reside and
cluster in the $L_o$ domain in their bound
states.\cite{huang1999nerve,bodosa2020preferential} Moreover, $L_o$
domains also promote the formation of amyloid-$\beta$ and fibril
aggregation, associated with the development of Alzheimer\rq s
disease.\cite{bodosa2020preferential,kedia2020real,colin2016membrane}
Viral assembly sites of HIV are generally located in the ordered
domains of the lipid membrane because of the affinity of gag protein,
a significant player in the viral assembly process, toward cholesterol
and sphingolipid.\cite{sengupta2019lipid} On the other hand, the
fusion peptide (FP) of HIV gp41 envelop protein exhibits no specific
preference to either ordered or disordered
domains.\cite{yang2017hiv,yang2016line}

The partition of peptides and proteins between membrane domains has
been also studied via computer simulations. For example, model
peptides, KALP and WALP, prefer to partition to the $L_d$ phase,
because of the lower free energy in the disordered
region.\cite{schafer2011lipid,kaiser2011lateral} Similar partitioning
preference toward $L_d$ regions was noted for rhodopsin, a G-protein
coupled receptor, and 7-TM protein
bacteriorhodopsin.\cite{bennett2013computer,periole2007g,DOMANSKI2012984}
CG simulations demonstrated that H-Ras proteins accumulate at
$L_d-L_o$ interfaces, while Hedgehog proteins prefer the $L_o$
domains.\cite{de2013molecular,janosi2012organization}

The affinity of proteins to different phases of lipid membranes
depends on many factors. Proteins with longer hydrophobic
transmembrane domain (TMD) tend to reside in the thicker liquid
ordered phase because of hydrophobic mismatch
considerations.\cite{lorent2020plasma,lin2018protein,bodosa2020} On
the other hand, proteins that have TMD with a larger accessible
surface area, exhibit lesser affinity to the $L_o$
phase.\cite{bodosa2020preferential,lorent2017structural} Chemical
modifications can change the affinity of proteins. For example, HIV
gp160, and N-RAS proteins exhibit affinity to ordered domains after
palmitoylation, whereas the distribution of palmitoylated tLAT remains
controversial.\cite{lin2018protein,levental2010greasing}

Proteins are not only attracted to different liquid phases of
heterogeneous lipid membranes, but also influence the phase behavior
itself. Some proteins like lectins can bind to carbohydrates/glycols
attached to lipids, leading to heterogeneous lateral organization of
lipids.\cite{NILSSON20071} BCRs are known to control the size and
stability of liquid ordered domains.\cite{stone2017protein} Specific
and non-specific interactions between TMDs of integral proteins and
lipids
\cite{cebecauer2018membrane,marsh2008protein,niemela2010membrane} and
hydrophobic mismatch \cite{mouritsen1984mattress} may also promote
formation of lipid domains. It is worth reminding here that another
protein component of the cell cortex, namely actin, which is also
believed to contribute to lipid raft
formation.\cite{kusumi2005paradigm} In general, the mechanisms by
which proteins influence the heterogeneity of the lipid membranes are
far from
clear.\cite{sperotto1989theory,hinderliter2001domain,hoferer2019protein}

The present study aims to extend the previously developed minimal
lattice model of ternary mixtures of saturated and unsaturated lipids
with Chol.\cite{PhysRevResearch.3.L042030,D2SM01025A,sarkar2023} Here,
we add to the model a small fraction of objects representing small
proteins (peptides), and examine their partition between the
liquid-disordered and liquid-ordered regions, and their influence on
the phase behavior of the mixture. As before, the model involves only
nearest-neighbor interactions. We keep the number of interaction
parameters minimal by assuming that the proteins have no direct
interactions between themselves, and considering only interactions
between the peptides and the saturated ordered chains. We analyze the
partition of the model proteins between the liquid phases, and explore
their influence on the formation, stability, and characteristics of
the liquid-ordered domains.

\section{Methods}

The lattice model introduced herein is an extension of a previous
model of ternary mixtures consisting of saturated (DPPC) and
unsaturated (DOPC) lipids with
Chol\cite{PhysRevResearch.3.L042030,D2SM01025A}. As in previous
studies, the simulations are conducted on a triangular lattice of
$121\times140=16940$ sites with periodic boundary conditions and
lattice spacing $l\simeq 0.56$ nm. This value of $l$ is 
set to match area density of DPPC in the liquid-ordered state 
(see details in ref.~\onlinecite{PhysRevResearch.3.L042030}). 
The lipids are modeled as dimers
with their two acyl chains occupying adjacent lattice sites, and Chol
is modeled as a monomer. Into this mixture, we now introduce small
proteins (peptides) that are represented as triangle-shaped trimers.

To sample the phase space of the system, we perform MC simulations
involving displacement of monomers, rotation of dimers (displacement
of one chain), and flips of trimers (reflection of one vertex across
the edge connecting the other two). Some lattice sites are left empty
to allow molecular diffusion within the system. Moves are accepted by
the Metropolis criterion, and only if the displaced particle lands on
a vacant site or a site occupied by a Chol monomer, in which case the
Chol swaps places with the displaced particle. The DOPC unsaturated
chains are disordered, while the DPPC saturated chains may be either
ordered or disordered. Therefore, the simulations also include
attempts to change the state of such chains. We define a MC time unit
as a collection of $1.05\times10^8$ trial moves, of which 95\% are
displacements/rotations/reflections of particles, and the rest 5\% are
attempts to change the state (ordered/disordered) of randomly chosen
DPPC chains. The length of the simulations extends between 1000 to
16000 time units, depending on the composition and the phase
behavior. Properties of interest have been calculated only after the
system equilibrated from the initial random configuration and the
energy saturated to the equilibrium value. For each simulated mixture,
we have performed independent runs with different initial
configurations to verify the consistency of the equilibrated states.

\begin{figure*}
\includegraphics[width=0.9\textwidth]{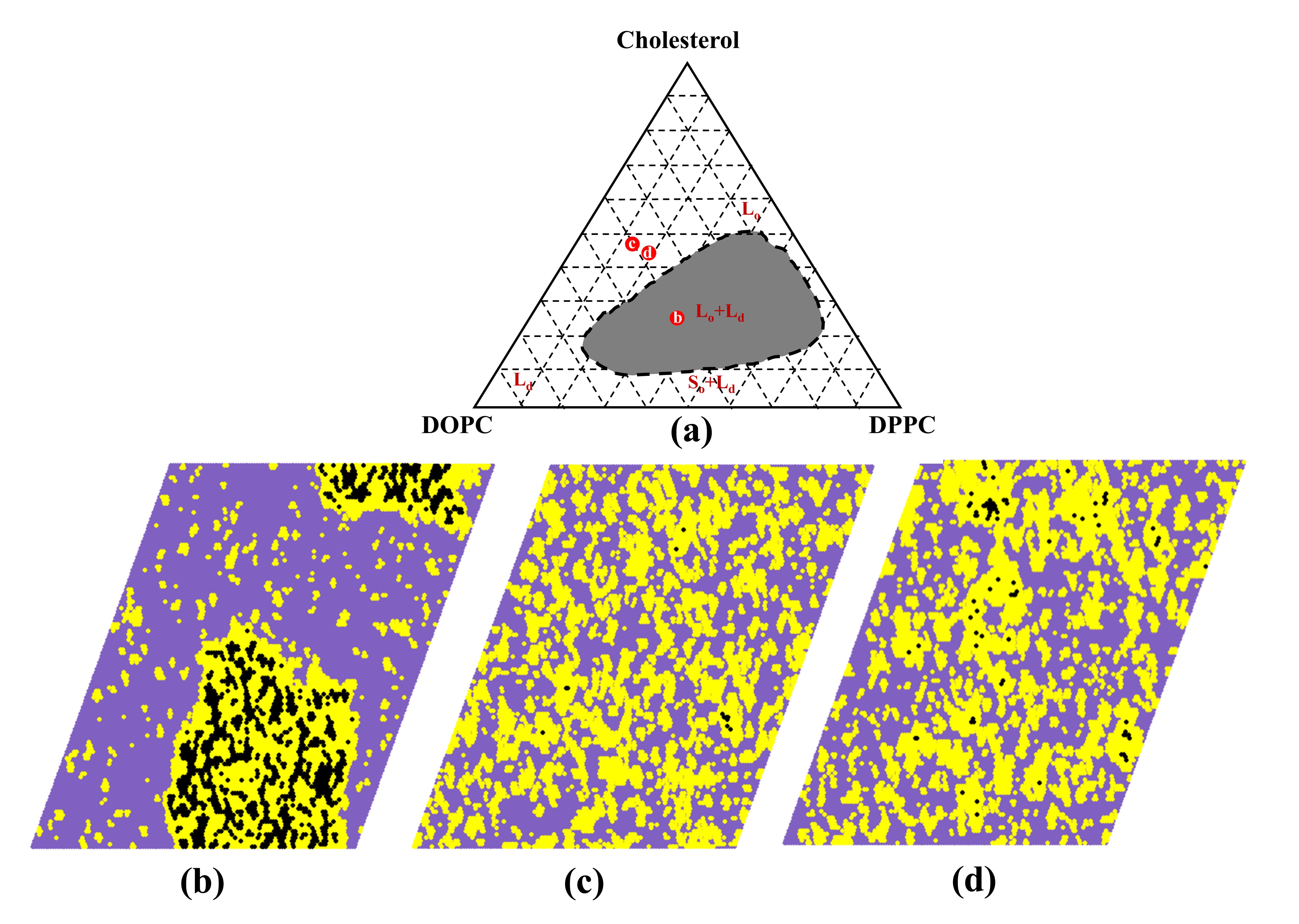}
 \centering
 \caption{(a) The phase diagram of a ternary DPPC/DOPC/Chol mixture at
   $T=298K$, adapted from ref. \onlinecite{veatch2004liquid}. The grey
   shaded area is the region of liquid-liquid phase coexistence. The
   red dots indicate the compositions of the simulated systems, whose
   equilibrium snapshots are shown in (b)-(d), respectively. The
   liquid-disordered ($L_d$), liquid-ordered ($L_o$), and gel ($S_o$)
   sites are colored in purple, yellow, and black, respectively. The
   compositions of the simulated systems are given in Table
   \ref{table:comp}.  }
 \label{fig:phasedia}
 \end{figure*}

To summarize, each lattice site can be assigned one of the
following six states:
\begin{inparaenum}[(i)]
\item void ($s=0$),
\item disordered DPPC ($s=1$),
\item ordered DPPC ($s=2$),
\item Chol\@($s=3$),
\item DOPC ($s=4$), and
\item protein ($s=5$).
\end{inparaenum}
In the model, the molecular forces are represented by nearest-neighbor
interactions only. Explicitly, the model Hamiltonian reads:
\begin{equation}
\begin{split}
E =&-\Omega k_BT \sum_{i}\delta_{s_i,1} -\sum_{i,j}\varepsilon_{22}\delta_{s_i,2}\delta_{s_j,2}\\
&-\sum_{i,j}\varepsilon_{23}\left[\delta_{s_i,2}\delta_{s_j,3}+\delta_{s_i,3}\delta_{s_j,2}\right]\\
&-\sum_{i,j}\varepsilon_{24}\left[\delta_{s_i,2}\delta_{s_j,4}+\delta_{s_i,4}\delta_{s_j,2}\right]\\
&-\sum_{i,j}\varepsilon_{25}\left[\delta_{s_i,2}\delta_{s_j,5}+\delta_{s_i,5}\delta_{s_j,2}\right]\\
&-\sum_{i,j}\varepsilon_{55}\delta_{s_i,5}\delta_{s_j,5},
\end{split}
\label{eq:hamil}
\end{equation}
where the sums run over all pairs of nearest neighbor sites $(i,j)$,
except for the first term, where the summation is over all the lattice
sites. This term accounts for the entropy gained when a DPPC chain is
in disordered ($s=1$), relative to the ordered ($s=2$) state. The
other terms in Eq.~(\ref{eq:hamil}) correspond, in order of
appearance, to the short-range attraction between an ordered DPPC
chain with another ordered DPPC chain, Chol, a disordered DOPC chain,
and a peptide. The last term represents protein-protein
interactions. We use the same model parameters as in our previous
studies: 
%of ternary mixtures
$\Omega=3.9$,
$\varepsilon_{22}=1.3\varepsilon$, $\varepsilon_{23}=0.72\varepsilon$. 
These values were set in the simulations of binary DPPC/Chol mixtures\cite{PhysRevResearch.3.L042030}
to reproduce the phase diagram of such mixtures as a function of the Chol mole fraction, at temperatures near the melting temperature of pure DPPC membranes $T_m=314K=0.9/k_B\varepsilon$.
%We set the relationship between the energy unit and the temperature of the simulations to $\varepsilon=k_BT/0.9$. 
In the following study where DOPC lipids were added to the system\cite{D2SM01025A}, we found that further
setting $\varepsilon_{24}=0$ yields results that  
%For this set of parameters,
%the phase diagram of the mixture as a function of the molar fractions
%of the components, 
agrees very well with the
experimentally-established phase diagram of DPPC/DOPC/Chol mixtures at
temperatures in the range between 280 and 300 degrees
Kelvin.\cite{veatch2007critical,veatch2005miscibility,veatch2003separation,veatch2004liquid}
The phase diagram of the mixture is plotted in
fig.~\ref{fig:phasedia}(a).  Importantly, it includes a two-phase
region of coexistence between the liquid-ordered and liquid-disordered
phases which is marked by the grey-shaded area. DPPC/DOPC/Chol is
known as a ``Type II'' mixture, i.e., a mixture where the two liquid
phases are macroscopically (thermodynamically) separated. 
%This feature of the liquid-liquid two-phase region is accomplished in the
%simulations by setting $\varepsilon_{24}=0$.

In this paper, we study the influence of a small density of peptides
on the phase diagram of the ternary mixture. For this purpose, we
include 100 peptide trimers in the mixture (covering 2\% of the lattice area), 
and simulate the system at different lipid-Chol compositions corresponding to different regions
of the phase space. Our aim here is {\em not}\ to investigate specific peptides, but to explore to possible impact of adding peptides to the DPPC/DOPC/Chol mixture. Since the peptide-DPPC attraction competes with the DPPC-DPPC attraction (which is the driving the formation of ordered domains), we chose to vary $\varepsilon_{25}$ between 0 and $2\varepsilon_{22}=2.6\varepsilon$ which, as will be shown below, covers a spectrum of distinct phase behaviors.
%The model parameter $\varepsilon_{25}$ representing the affinity of the proteins to ordered DPPC chains is varied between 0 to $2.6\varepsilon$ (twice the affinity parameter $\varepsilon_{22}$ of the chains to each other). 
We also set $\varepsilon_{55}=0$ and so the proteins in the present study are
assumed to be non-interacting.

\begin{table*}
\centering
\caption{Compositions of ternary lipid mixtures simulated in this
  work. The roman alphabets within the brackets in the first column
  indicates the corresponding snapshot in Fig. \ref{fig:phasedia}}
\label{table:comp}
\begin{tabular}{|c|c|c|c|}
\hline
\textbf{Designation} &\textbf{DPPC} & \textbf{DOPC} & \textbf{Chol}\\
{} & \textbf{(mol\%)} & \textbf{(mol\%)} & \textbf{(mol\%)}\\
\hline 
35DPPC (b) & 35 & 40 & 25\\
\hline 
12DPPC (c) & 12 & 40 & 48\\
\hline
18DPPC (d) & 18 & 38 & 44\\
\hline
\end{tabular}
\end{table*}

In the following section, we show many equilibrium snapshots of
mixtures at different compositions and for different values of the
model parameter $\varepsilon_{25}$. In these snapshots, the system is
divided into liquid-disordered ($L_d$), liquid-ordered ($L_o$), and
gel ($S_o$) regions, which are displayed in purple, yellow, and black,
respectively. The division of the system between the different regions
is based on the simple algorithm introduced in our previous simulation
study of ternary mixtures, which has been modified here to account for
the presence of proteins. In the original algorithm (see full details
in ref.~\onlinecite{D2SM01025A}), we assign each state of the lattice
sites with the following orderliness score, $S_i$: A void, ordered
DPPC chain, disordered DPPC chain, Chol, and DOPC chain, gets a score
of 0, 2, -0.5, 1, and -1, respectively. The order parameter, $G_i$, of
a lattice site is given by
\begin{equation}
  G_i= Sc_i+W_i\sum_{j=1}^6{Sc_j},
  \label{eq:score}
\end{equation}
where the sum runs over the six nearest-neighbor sites. In ternary
mixture simulations without proteins, we set $W_i=1$. In the present
study, a site with a protein gets a zero score, and $W_i=6/(6-N^p_i)$,
where $0\leq N^p_i< 6$ is the number of nearest-neighbor sites
occupied by proteins. If all the neighbors are occupied by proteins
($N^p_i=6$, a scenario that actually never encountered in the
simulations), then we simply set $G_i=Sc_i$. Sites with negative
(non-negative) grades, $G_i<0$ ($G_i\geq 0$) are associated with the
liquid-disordered (liquid-ordered) regions. The site with the maximum
possible order parameter, $G_i=14$, constitute the gel region. The
sites hosting the proteins are plotted in green irrespective of their
grade.

%We have computed the histogram of distribution of $G_i$ based on its average over 100 frames, after the system equilibrated properly.
%We have also calculated the percentage of proteins completely immersed into the ordered phase ($\phi$). A triangular protein is completely immersed into the ordered phase only if all of its nine nearest neighbours belong to $L_o$ or $S_o$ phase.

Some properties of the larger ordered domains have been computed.  The
Hoshen–Kopelman algorithm\cite{hoshen1976percolation}, adopted for a
periodic triangular lattice, has been used to identify these domains,
and their radius of gyration was also calculated
\begin{equation}
  R_g=\sqrt{\frac{1}{2N^2}\sum_{i\neq j}|r_i-r_j|^2},
\end{equation}
where the sum is carried over all the pairs of points belonging to a domain.

\section{Results and Discussion}

Our goal is to study how the addition of a small fraction of peptides
modifies the lateral organization and phase behavior of ternary
mixtures of saturated DPPC and unsaturated DOPC lipids with Chol. For
this purpose, we add 100 trimers representing small peptides to
mixtures with lipid compositions given in Table \ref{table:comp}. In
this table, point (b) corresponds to a mixture exhibiting macroscopic
$L_d+L_o$ coexistence.  The other two compositions [(c) and (d)]
correspond to the one-phase region where the system is homogeneous on
macroscopic scales, and features local density fluctuations appearing
as small nanoscopic ordered regions in a liquid-disordered
background. Point (d) is close to a miscibility transition point, as
evident from the larger size of the ordered domains and the fact that
the system is on the verge of percolation.

\begin{figure*}
\centering
\includegraphics[width=1.0\textwidth]{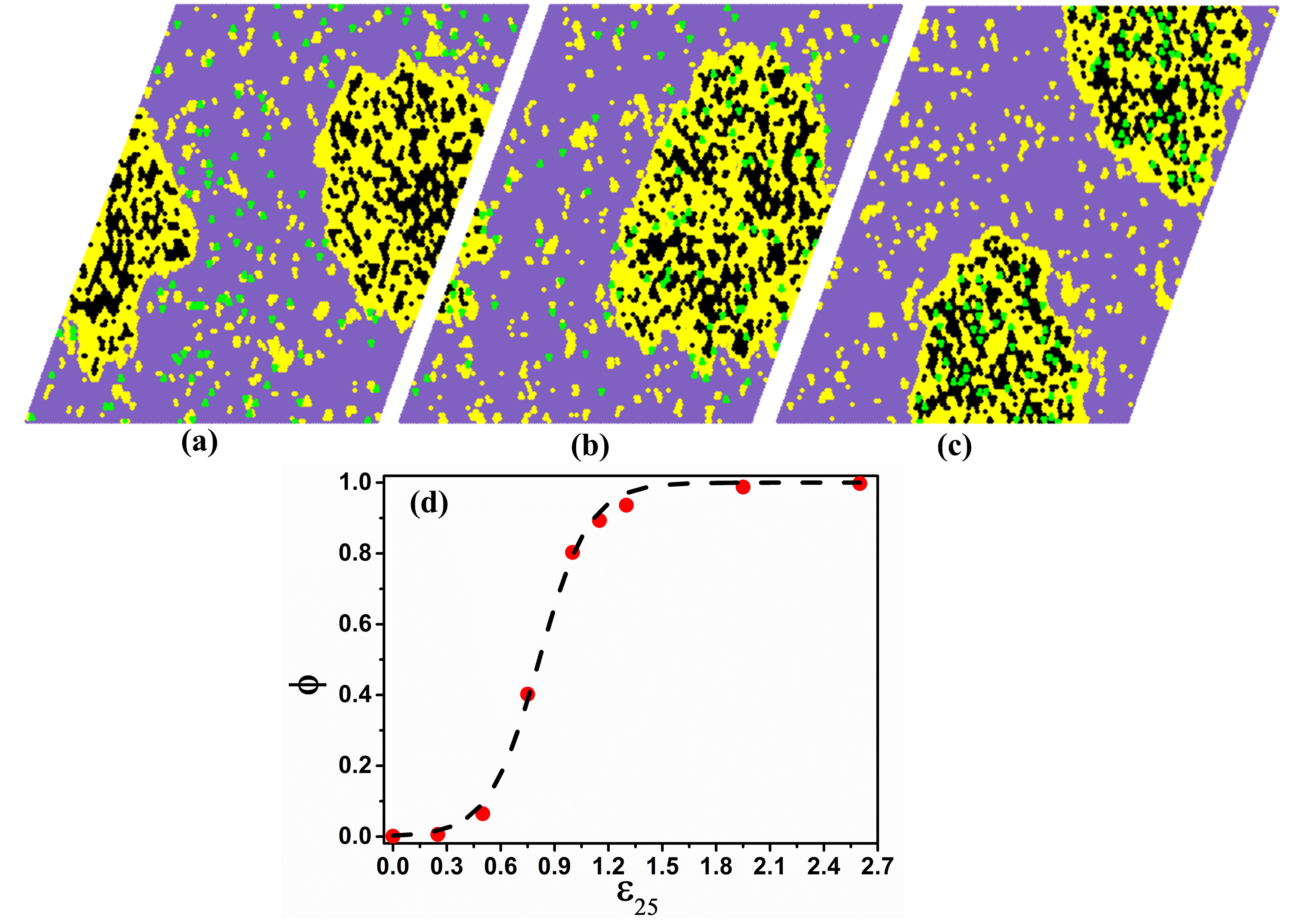}
\caption{Partitioning of proteins in the two phase region (35DPPC) of
  DPPC/DOPC/Chol mixture for varying strength of
  $\varepsilon_{25}$. Snapshots (a), (b), and (c) show equilibrium
  distributions of systems corresponding to $\varepsilon_{25}=0$,
  $0.75\varepsilon$, and $1.95\varepsilon$, respectively. Color coding
  the same as in fig.\,\ref{fig:phasedia}, with proteins marked by
  green. (d) The fraction, $0\leq\phi\leq 1$ of the proteins
  completely inside the $L_o$ phase, as a function of
  $\varepsilon_{25}$. The dashed line is a fit of the results to
  Eq.~(\ref{eq:partition}) with $a\simeq 7.2$ and $b\simeq 5.8$.}
\label{fig:2phase}
\end{figure*}

\subsection{Protein partitioning in the two-phase region}

We start with the simpler case of the two-phase regime. As discussed
earlier, different biological processes involve protein accumulations
in either the liquid-disordered or liquid-ordered regions. This
feature is easily captured in our simulations, as illustrated in
fig.~\ref{fig:2phase}, showing how the affinity of the proteins to the
two phases varies as a function of the interaction parameter
$\varepsilon_{25}$. The snapshots in the figure show equilibrium
distributions for 35DPPC [point (b), in fig.~\ref{fig:phasedia}],
corresponding to (a) $\varepsilon_{25}=0$, (b) $0.75\varepsilon$, and
(c) $1.95\varepsilon$. The trend in these snapshots is clear. The
$L_o$ phase is stabilized by the short-range packing attraction of the
saturated ordered lipids to the Chol molecules ($\varepsilon_{23}$)
and, especially, to each other ($\varepsilon_{22}$). Therefore,
insertion of proteins into the liquid-ordered phase depends on the
competition between these packing interactions and the attraction of
the proteins to the ordered chain ($\varepsilon_{25}$). Accordingly,
we see in fig.~\ref{fig:2phase} that proteins are completely expelled
from the $L_0$ phase for $\varepsilon_{25}=0$ (a), partially penetrate
the ordered region for $\varepsilon_{25}=0.75\varepsilon$ (b), and
become fully encapsulated in it for $\varepsilon_{25}=1.95\varepsilon$
(c).

A quantitative representation of the protein partitioning between the
liquid-disordered and liquid-ordered for 35DPPC mixture is found in
Fig. \ref{fig:2phase}(d), where the fraction, $0\leq\phi\leq 1$ of the
proteins completely inside the $L_o$ phase (protein trimers having all
of their 9 neighbors from the $L_o$ phase) is plotted as a function of
$\varepsilon_{25}$. The graph shows a sharp crossover in the
distribution of the proteins between the two phases. The change in the
protein distribution can be related to the difference in the solvation
free energy, $\Delta F$, of the small proteins in the two liquid
phases. It is reasonable to assume that $\Delta
F=-a\varepsilon_{25}+b$, where the first term represents the
attraction between the proteins and the ordered DPPC chains, with
$a\leq 12$ being the average number of nearest neighbors interactions,
per protein, with ordered DPPC chains. The second term accounts for
all the other thermodynamic considerations, including the competing
lipid-lipid and lipid-Chol interactions in the $L_o$ phase, as well as
the contribution due to mixing entropy. At a small density of
proteins, collective effects in the partitioning behavior may be
neglected, and we can expect that
\begin{equation}
  \phi=\frac{\exp{[(a\varepsilon_{25}-b)/k_BT]}}
            {1+\exp{[(a\varepsilon_{25}-b)/k_BT]}}
  \label{eq:partition}
\end{equation}
The dashed curve in fig.~\ref{fig:2phase}(d) is a fit of the results
to Eq.~(\ref{eq:partition}) with $a\simeq 7.2$ and $b\simeq 5.8$.

\begin{figure*}
\centering
\includegraphics[width=1.0\textwidth]{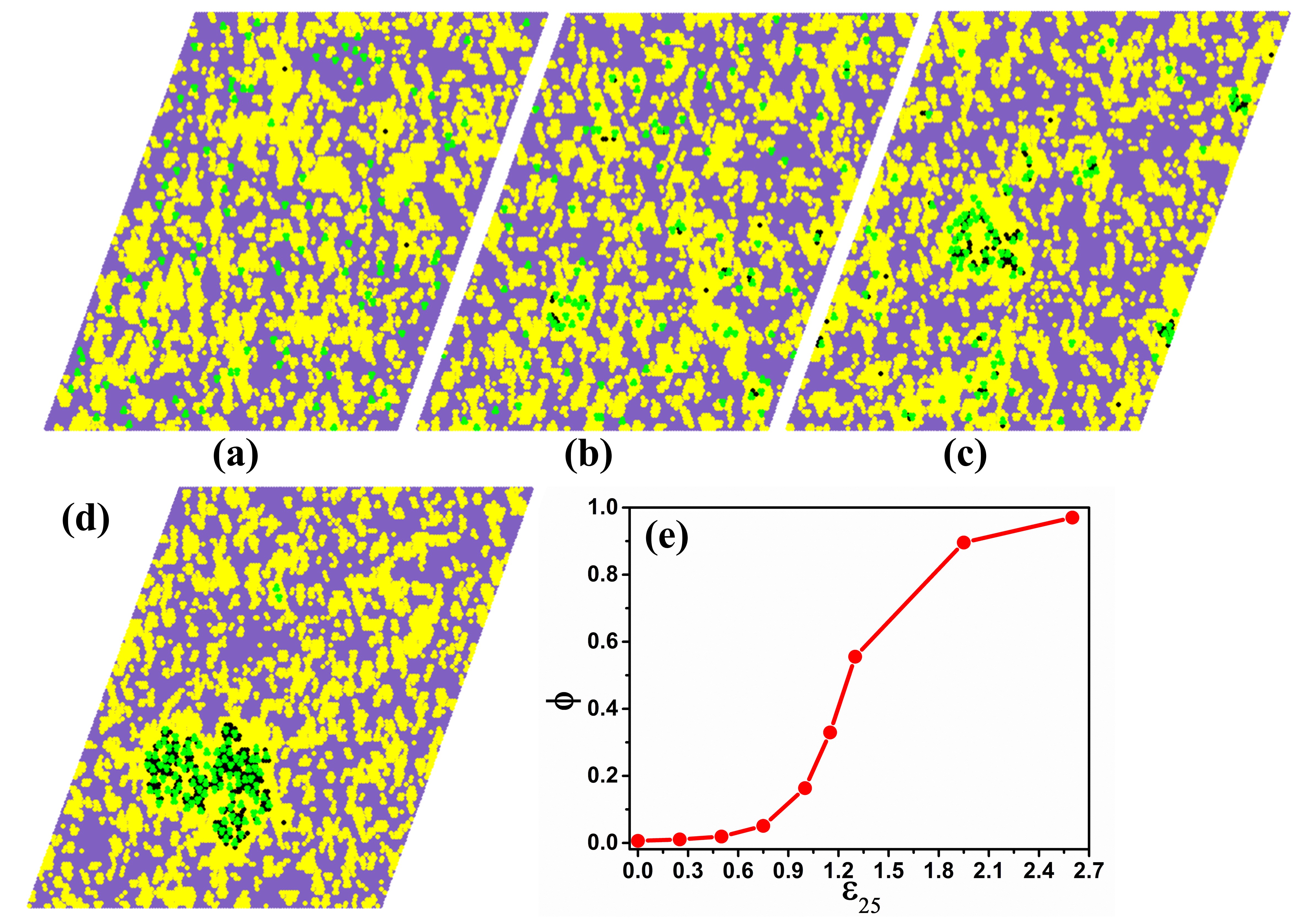}
\caption{Partitioning of proteins in DPPC/DOPC/Chol mixture in
  one-phase region (12DPPC) for varying strength of
  $\varepsilon_{25}$. Snapshots (a)-(d) show equilibrium distributions
  of systems corresponding to $\varepsilon_{25}=0$, $1.0\varepsilon$,
  $1.3\varepsilon$, and $1.95\varepsilon$, respectively. Color coding
  the same as in fig.~\ref{fig:2phase}. (e) The fraction,
  $0\leq\phi\leq 1$ of the proteins completely inside the $L_o$ phase,
  as a function of $\varepsilon_{25}$. The line is a guide to the
  eye.}
\label{Fig:12dppc}
\end{figure*}

\subsection{Formation of large domains in the one-phase region}

The influence of small proteins on mixtures in the one-phase region is
more interesting than in the two phase region. The one-phase region is
locally inhomogeneous, featuring nanoscopic ordered domains. Addition
of even a small amount of proteins can change this local organization
dramatically and, as shown below, may lead to the formation of much
larger domains. This is interesting in light of the question lingering
about the formation and sizes of liquid-ordered domains in model
mixtures at physiological temperature 37C, and the role played by the
proteins in the assembly and growth of raft domains in biological
membranes.\cite{kusumi2020defining,rosetti2017sizes}

Computational results from the simulations of the one-phase 12DPPC
mixture are displayed in fig.~\ref{Fig:12dppc}. The trend in the
partitioning behavior of proteins is quite similar to the behavior in
fig.~\ref{fig:2phase} of the macroscopically phase-separated 35DPPC
mixture. The fraction of proteins in the liquid-ordered domains
increases with the interaction free energy energy, $\varepsilon_{25}$,
between them and ordered DPPC chains. This is evident from the
sequence of equilibrium snapshots fig.~\ref{Fig:12dppc}(a)-(d),
corresponding to $\varepsilon_{25}=0$, $1.0\varepsilon$,
$1.3\varepsilon$, and $1.95\varepsilon$, respectively, as well as from
fig.~\ref{Fig:12dppc}(e) showing the fraction of proteins fully
residing in the liquid-ordered domains. Notice that, in comparison to
the two-phase region (fig.~\ref{fig:2phase}), the transition of the
proteins from the disordered to the ordered regions in the one-phase
system occurs at higher values of $\varepsilon_{25}$, which is
anticipated since the fraction of saturated DPPC lipids is
smaller. Also, because the system separates locally rather than
macroscopically, there is a larger interfacial contact line between
the liquid phases and, therefore, a higher fraction of proteins reside
between them rather than inside the liquid-ordered regions.

%However, for $\varepsilon_{25}=2$\varepsilon_{22}=2.6\varepsilon$, we observe [see fig.~\ref{Fig:12dppc}(d)] that almost all the proteins assemble in a single liquid-ordered domain that is much larger than the rest of the smaller liquid-ordered regions. 

\begin{figure*}
\centering
\includegraphics[width=1.0\textwidth]{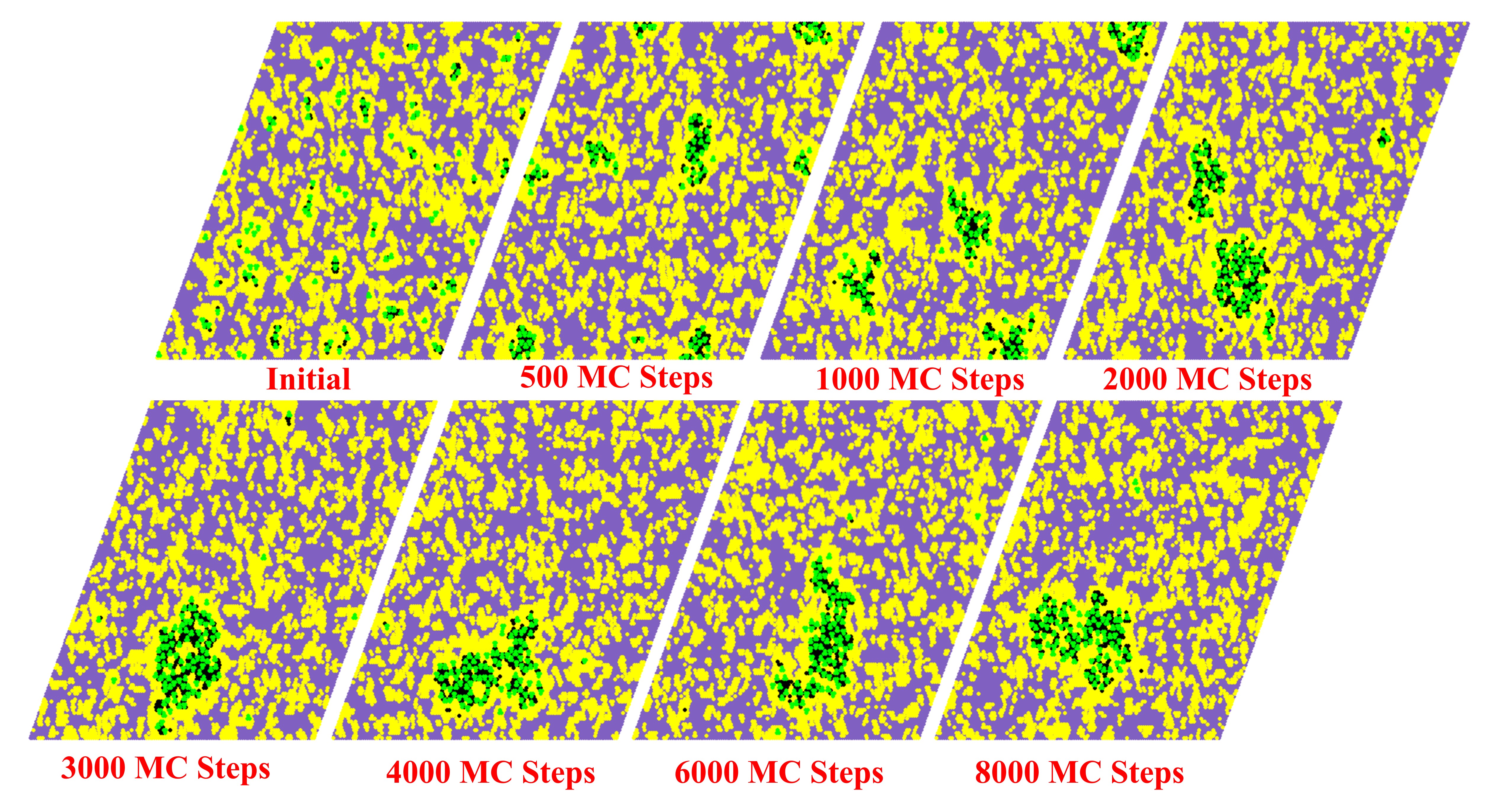}
\caption{A sequence showing the temporal evolution and the formation
  of a large cluster containing almost all the proteins in a 12DPPC
  mixture with $\varepsilon_{25}=1.95\varepsilon$. Color coding as in
  fig. \ref{fig:2phase}.}
\label{fig:clusdyn} 
\end{figure*}

\begin{figure*}
\centering
\includegraphics[width=0.9\textwidth]{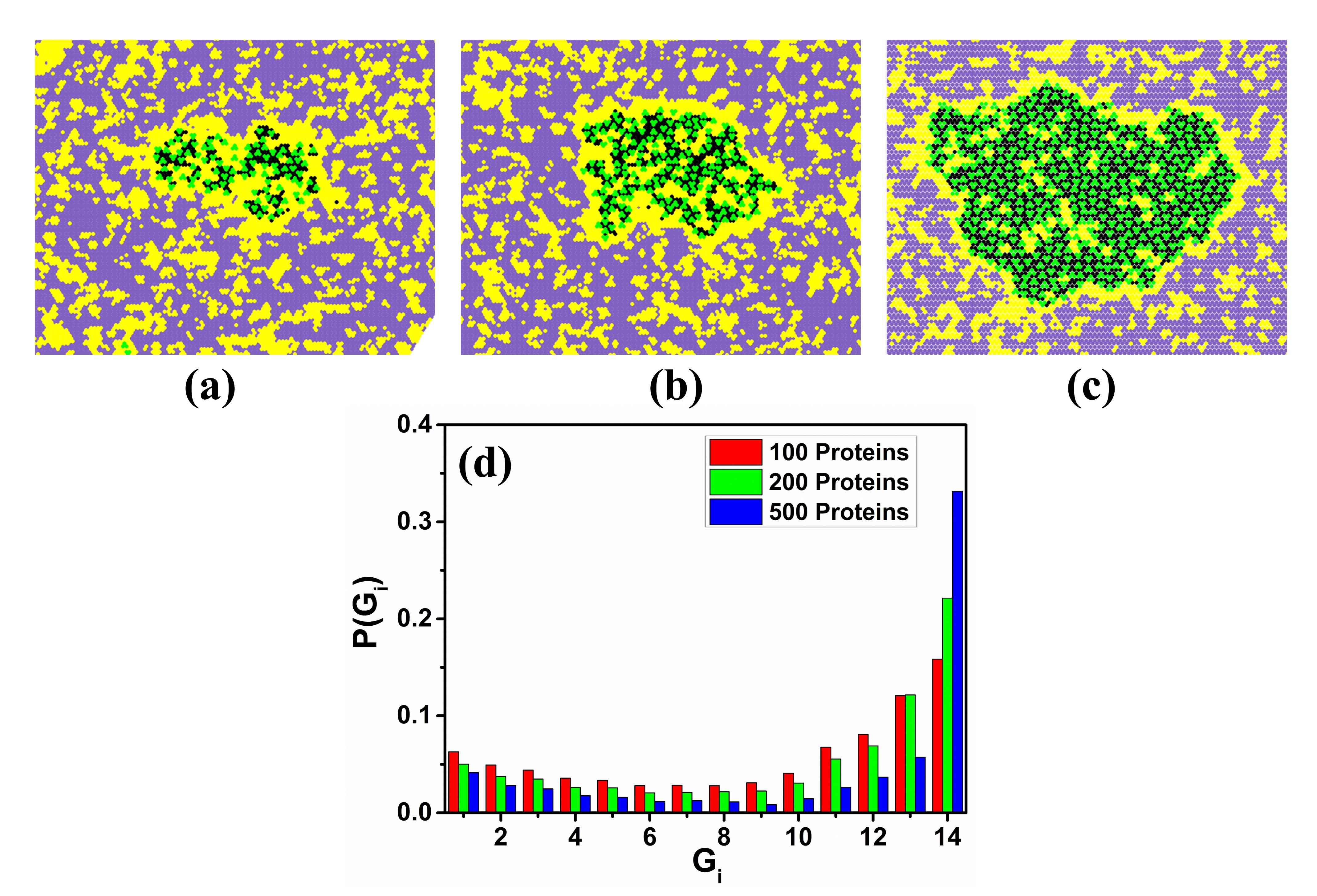}
\caption{Equilibrium snapshots showing the large ordered domain formed
  in 12DPPC mixtures with $\varepsilon_{25}=1.95\varepsilon$ and (a)
  100, (b) 200, and (c) 500 proteins. (d)The distribution histogram of
  the values of order parameter $G_i$ of the sites belonging to the
  large ordered domain.}
\label{fig:score12dppc}
\end{figure*}

An interesting observation is that, together with the gradual
migration of the proteins into the liquid-ordered domains at higher
values of $\varepsilon_{25}$, larger domains surrounding the proteins
begin to appear in the mixture. This trend can be observed in the
snapshot in fig.~\ref{Fig:12dppc}(c) corresponding to
$\varepsilon_{25}=1.3\varepsilon$, showing several larger
protein-containing domains. By following the dynamics of the system
(see Supplementary Material, SI movie 12DPPC\_1.3.mp4), it can be
concluded that these are metastable dynamic domains: Their shapes and
sizes constantly change as they form, sometimes merge with other
domains, and eventually disintegrate.

Larger and more stable domains are obtained for higher values of
$\varepsilon_{25}=1.95\varepsilon$, see fig.~\ref{Fig:12dppc}(d). The
sequence of snapshots in fig.~\ref{fig:clusdyn} show how a single
large domain that contains almost all the proteins dispersed in the
system, evolves via the coalescence of smaller domains (see
Supplementary Material, SI movie 12DPPC\_1.95.mp4). This observation
marks the fundamental difference between the two-phase 35DPPC mixture
discussed above, and the one-phase 12DPPC mixture. In both cases the
proteins show affinity to the DPPC-rich ordered domains, but in the
latter case they also change the lipid distribution in the
mixture. The formation of a single large domain suggests that the
addition of a small density of proteins with strong affinity to the
saturated lipids can induce phase separation between a $L_o$ phase
that is rich in proteins and saturated lipids, and a $L_d$ phase with
nanometric ordered domains, which is DPPC-poor and completely depleted
of proteins. The fact that, despite a clear loss of mixing entropy,
the proteins are not dispersed in the many liquid-ordered nanometric
domains but aggregate in a single larger domain, is a clear indication
that the mixture undergoes a phase transition that is driven by the
attractive interaction between the proteins and the ordered saturated
chains.

The size of the large domain is limited by the amount of the available
proteins, as can be seen in the snapshots in
fig.~\ref{fig:score12dppc}, showing equilibrium configurations of
12DPPC mixtures with $\varepsilon_{25}=1.95\varepsilon$ and with (a)
100, (b) 200, and (c) 500 proteins, respectively. The growth in the
size of the large domain is not only due to the addition of proteins
to the mixture, but also because of the recruitment of DPPC
lipids. About 40\% of them are found in the large ordered domain in
(a), and this fraction grows to $\approx 60$\% and $\approx 90$\% in
(b) and (c), respectively. In contrast , the partition of the other
components between the large domain and the surrounding does not
change with the addition of proteins, and we find that less than 3\%
of the DOPC lipids and $\approx 13$\% of the Chol molecules are in the
ordered domain. Thus, the composition {\em within}\/ the large domain
is different in snapshots fig.~\ref{fig:score12dppc} (a)-(c), implying
that the phase of the domain may be also different. One of the
hallmarks of the $L_o$ phase is the existence of gel-like
nano-clusters\cite{D2SM01025A,PhysRevResearch.3.L042030}, which in the
absence of proteins (i.e., for mixtures containing only lipids) are
composed of tightly packed saturated lipids. In mixtures also
containing proteins with strong affinity to saturated lipids, these
nano-clusters are nucleated around the proteins, as can be seen in
fig.~\ref{fig:score12dppc} (a) by the overlap between the locations of
of the proteins (marked by green color) and the gel-like regions
(marked by black). With the addition of more proteins to the system
and the corresponding changes in the composition of the large domain,
the phase changes gradually from $L_o$ (liquid-ordered) in
fig.~\ref{fig:score12dppc} (a) to $S_o$ (gel) in
fig.~\ref{fig:score12dppc} (c). This trend is summarized in
fig.~\ref{fig:score12dppc} (d), showing the distribution histograms of
the values of the order parameter $G_i$ [Eq.~(\ref{eq:score})] in the
large ordered domain. In all three cases corresponding to different
numbers of proteins, a significant fraction of the domain sites are
associated with the gel state ($G_i=14$). This number grows markedly
with the number of proteins, which indicates the transformation of the
ordered domain from a liquid-ordered to gel phase.

\begin{figure*}
\centering
\includegraphics[width=0.9\textwidth]{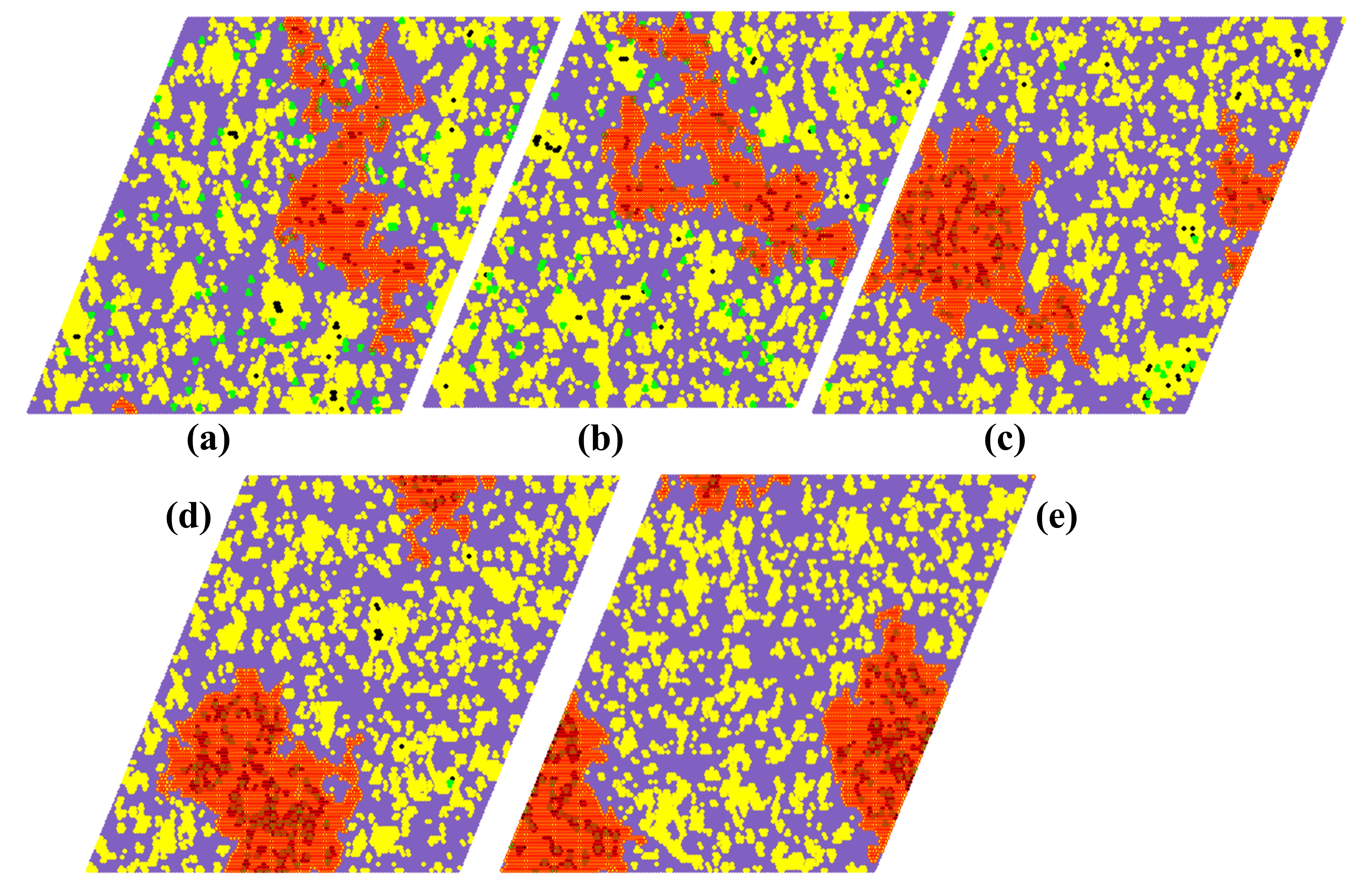}
\caption{Equilibrium configurations of 18DPPC mixture with
  $\varepsilon_{25}=0$ (a), $0.75\varepsilon$ (b), $1.3\varepsilon$
  (c), $1.95\varepsilon$ (d), and $2.6\varepsilon$ (e). The mixture
  contains 100 proteins. Color coding as in fig. \ref{fig:2phase},
  except for the largest ordered domain in the mixture, which is
  colored by red.}
\label{fig:score18dppc}
\end{figure*}

To summarize, proteins with sufficiently strong attraction to
saturated lipids (stronger than the attraction of the lipids to each
other) can drive the system to phase separate by serving as nucleation
centers for the formation of liquid-ordered domains. This mechanism
has been speculated by experimental
works,\cite{sezgin2017mystery,komura2016raft} and has been
demonstrated in previous ultra CG and lattice
simulations.\cite{hoferer2019protein} At lower affinities and low
densities of proteins, the system may still exhibit metastable dynamic
domains whose size may be comparable to lipid rafts in complex
biological membranes.

\begin{table*}
\centering
\caption{Fractions of proteins and DPPC lipids partitioned in the
  large ordered domain, and data on its gyration radius, $R_g$, in
  18DPPC mixtures, as a function of $\varepsilon_{25}$.}
\begin{tabular}{|c|c|c|l|l|}
\hline $\mathbf{\varepsilon_{25}}$ & \textbf{\%Protein} & \textbf{\%
  DPPC} & \textbf{$R_g$} (Mean) \textbf{(\r{A})} & \textbf{$R_g$} (SD)
\textbf{(\r{A})}\\ \hline 0.0 & $5.0 \pm 2.5$ & $23.0 \pm 7.0$ & 167 &
67 \\ \hline 0.75 & $14.5 \pm 6.0$ & $19.0 \pm 7.0$ & 229 & 93
\\ \hline 1.30 & $87.0 \pm 6.0$ & $40.0 \pm 4.5$ & 263 & 15 \\ \hline
1.95 & $98.5 \pm 1.0$ & $52.0 \pm 2.5$ & 358 & 8 \\ \hline
\end{tabular}
\label{18dppctab}
\end{table*}

\begin{figure*}
\centering
\includegraphics[width=0.8\textwidth]{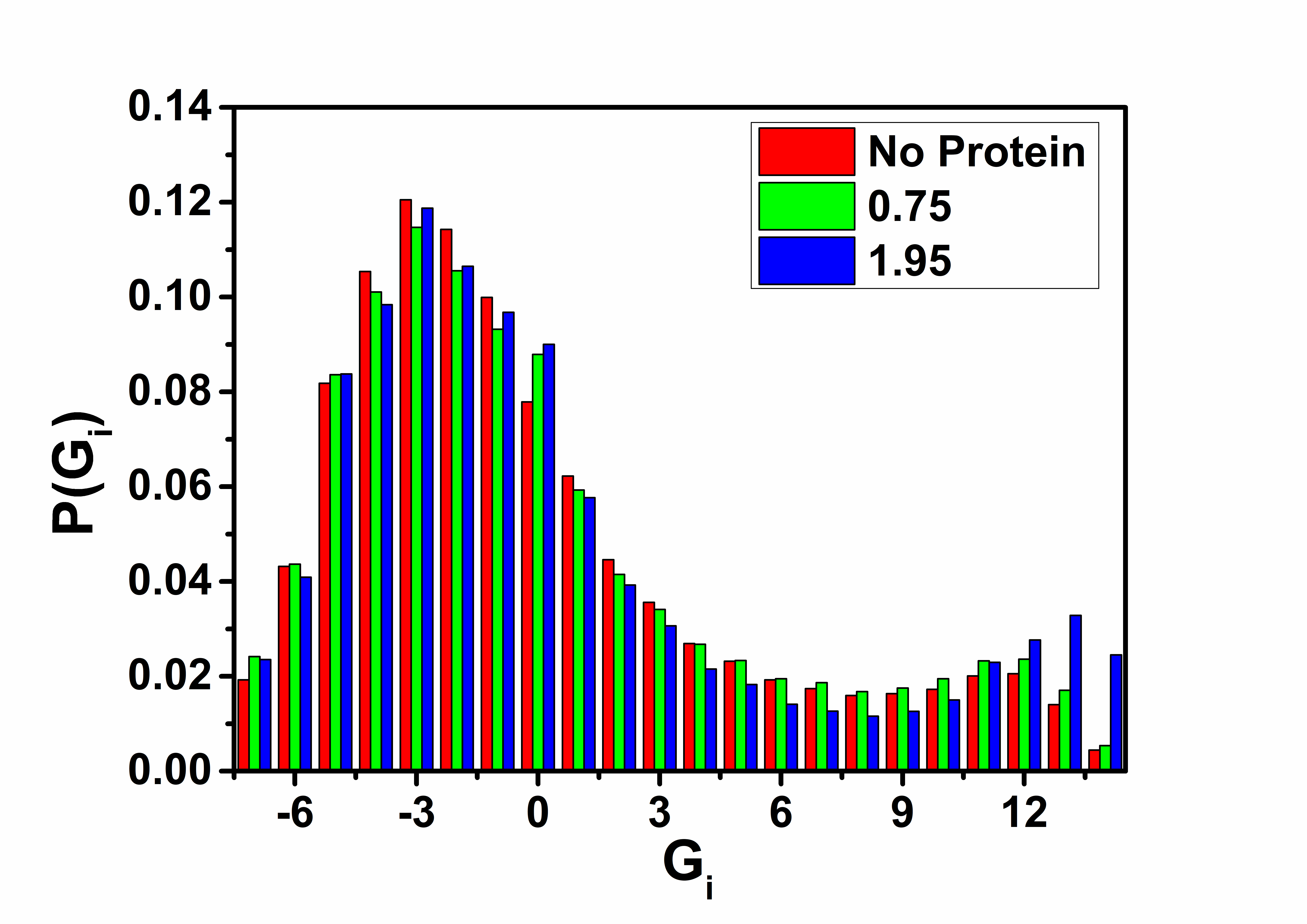}
\caption{The distribution histogram of the values of order parameter
  $G_i$ of 18DPPC mixtures with no proteins, 100 proteins with
  $\varepsilon_{25}=0.75\varepsilon$, 100 proteins with
  $\varepsilon_{25}=1.95\varepsilon$.}
\label{fig:larclus18dppc}
\end{figure*}

Fig.~\ref{fig:score18dppc} shows equilibrium configurations of the
18DPPC mixture, where the largest ordered cluster is colored by
red. The snapshots in (a)-(e) correspond to $\varepsilon_{25}=0$,
$0.75\varepsilon$, $1.3\varepsilon$, $1.95\varepsilon$, and
$2.6\varepsilon$, respectively. The main difference between the 18DPPC
mixture and the 12DPPC mixture is that the former is located close to
the phase boundary separating the one- and two-phase regions (see
fig.~\ref{fig:phasedia}). Therefore, the ordered domains appearing in
snapshots (a) and (b) are larger than those appearing for similar
values of $\varepsilon_{25}$ at 12DPPC. They look branched and
resemble percolation clusters that tend to form when mixtures are in
the vicinity of the the phase transition critical point. Other than
this, the system behaves quite similarly to the 12DPPC mixture: It
undergoes phase separation when the attraction of the proteins to the
DPPC lipids is stronger than the affinity of the lipids to each
other. The phase transition can be read from the distribution
histograms of the values of the order parameter $G_i$ shown in
fig.~\ref{fig:larclus18dppc}. These histograms are computed for the
entire mixture (not only for the largest ordered domain, as in
fig.~\ref{fig:score12dppc}) for 18DPPC mixture with no proteins, and
with 100 proteins with $\varepsilon_{25}=0.75\varepsilon$, and
$\varepsilon_{25}=1.95\varepsilon$. In the first two histograms, only
a tiny fraction of the system is identified as gel-like. In the third
case (100 proteins with $\varepsilon_{25}=1.95\varepsilon$), the
bimodal nature of the distribution is evident, with the second peak at
high values of $G_i$ (including a growing fraction of gel-like sites
with the maximum value $G_i=14$). As noted above, this is a
characteristic feature of the $L_o$ phase. More information about the
structural changes associated with the transition to two-phase
separation can be found in table~\ref{18dppctab}, where details about
the composition and structure of the largest ordered domain are
provided. The data highlights two major differences between the large
domains in the one-phase ($\varepsilon_{25}=0$ and $0.75\varepsilon$)
and the two-phase cases ($\varepsilon_{25}=1.3\varepsilon$ and
$1.95\varepsilon$). One is the fact that in the two-phase case, the
proteins are found in the $L_o$ large domain, whereas in the one-phase
region, they are distributed in the entire mixture. The $L_o$ domain
in the two-phase region also attracts a larger fraction of the
saturated DPPC lipids. These trends were also observed in the 12DPPC
mixture discussed above. The other notable observation from the table
is related to the size (gyration radius) of the largest domain. The
striking piece of information is not the steady growth in the size of
the domain with $\varepsilon_{25}$ [$R_g$ (mean)], but the standard
deviation [$R_g$ (SD)] which measures the characteristic fluctuations
in the domain size. The data reflects the stability of the $L_o$ phase
domain in the two-phase region where the size of the ordered domain is
nearly constant vs.~the dynamic nature of the ``percolation'' domain
in the one-phase region whose size varies considerably in time.

\section{Conclusions}
Ternary mixtures of saturated and unsaturated lipids with Chol serve
as minimal model systems for studying phase separation in complex
biological membranes. Under suitable conditions, they exhibit
coexistence of liquid-ordered domains with a liquid-disordered
matrix. This phase behavior is believed to be relevant to lipid rafts, 
which are dynamic liquid-ordered domains of typical size of several
tens of nanometers, that ``float'' on the cell surface.

\begin{figure*}
\centering
\includegraphics[width=0.8\textwidth]{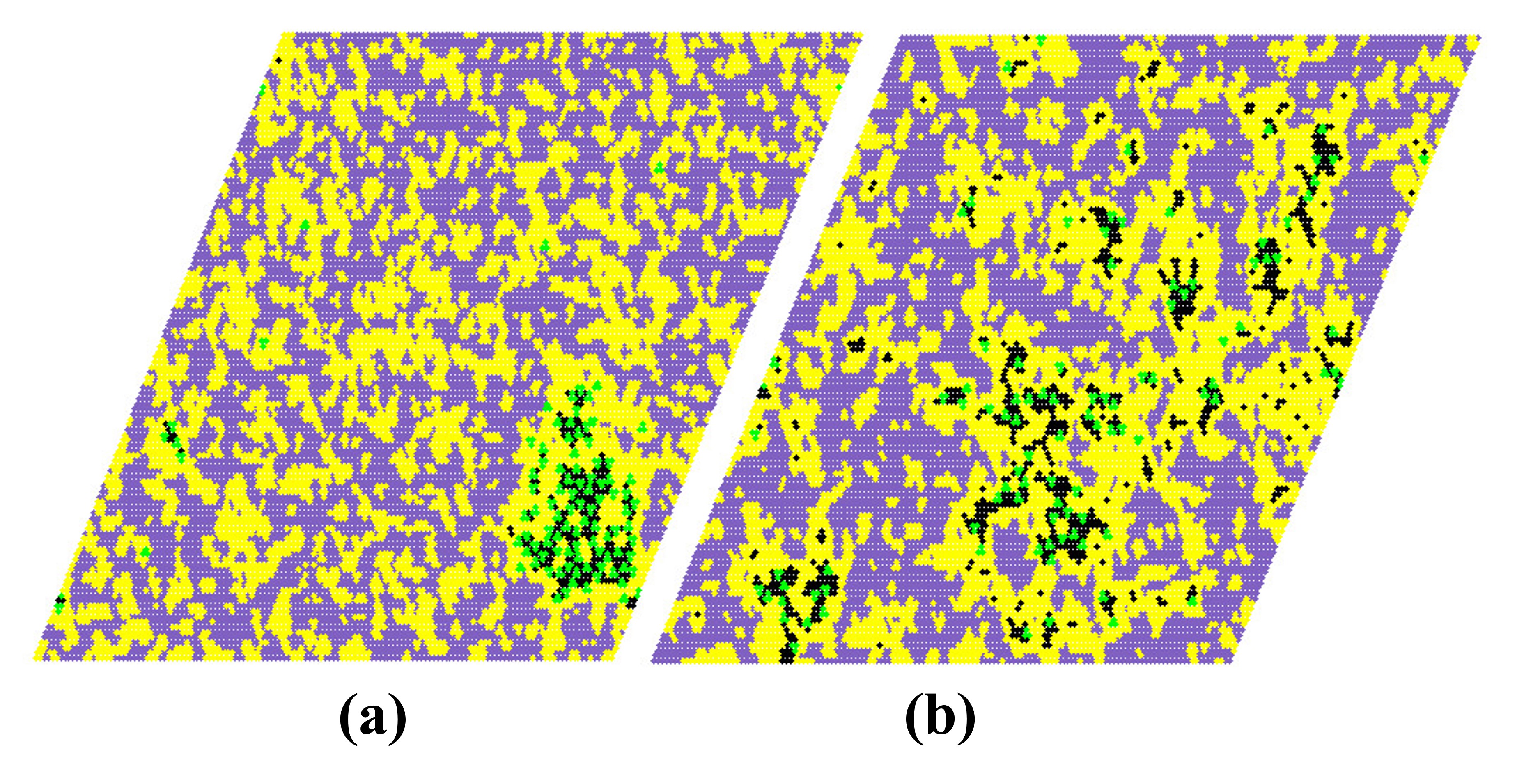}
\caption{Equilibrium snapshots of Type I mixtures with $\varepsilon_{24}=0.4\varepsilon$. (a) and (b) show 12DPPC and 35DPPC mixtures, respectively, In both cases, we set $\varepsilon_{25}=1.95\varepsilon$.} 
\label{fig:type1}
\end{figure*}

Lipid rafts contain certain proteins that are involved in different
biological processes. Here, we extended a previously published lattice
model of ternary DPPC/DOPC/Chol mixtures to include small proteins
(peptides). We perform extensive MC simulations to investigate two
closely-related phenomena - the impact of the proteins on the phase
behavior, and the distribution of the proteins between the liquid
phases. We focus on, so called, Type II mixtures, i.e., mixtures that
in the absence of proteins exhibit macroscopic liquid-liquid phase
separation. The proteins in our simulations do not interact directly
with each other. Thus, the extension of the model involves the
addition of only a single interaction parameter, $\varepsilon_{25}$,
corresponding to the affinity between the proteins and the saturated
ordered lipid chains. In future works we plan to extend our
investigations of Type I mixtures, which in our previous studies\cite{D2SM01025A,sarkar2023} were 
observed when the model parameter $\varepsilon_{24}$ was set to a value larger than $0.3\varepsilon$. 
Fig.~\ref{fig:type1} shows preliminary results from ongoing Type I mixture simulations for $\varepsilon_{24}=0.4\varepsilon$. Fig.~\ref{fig:type1}(a) is a 12DPPC type I mixture at $\varepsilon_{25}=1.95\varepsilon$, and is remarkably similar to fig.~\ref{Fig:12dppc}(d) which shows the type II mixture with the same model parameters except for the value of $\varepsilon_{24}$. In contrast, fig.~\ref{fig:type1}(b) displays a type I 35DPPC mixture, which looks markedly different than its type II counterpart shown in fig.~\ref{fig:2phase}(c). This difference may not be surprising considering that the main difference between Type I and Type II mixtures is in the two-phase region.

In addition to changing the Type of the mixture by variations of the model parameter 
$\varepsilon_{24}$, we also consider performing simulations with other model 
parameters like $\varepsilon_{35}$ (Chol-peptide) and $\varepsilon_{45}$ (DOPC-peptide), 
which should lead to an even richer phase behavior. A particularly 
interesting example is of mixtures where the peptides have no strong affinity to either 
of the liquid phases, but rather favor the proximity of the Chol which tend to be
present in both of them.

Not surprisingly, we find that the migration of proteins into the
liquid-ordered domains depends strongly on $\varepsilon_{25}$. More
precisely, since proteins embedded in liquid-ordered domains take the
place of saturated lipids and Chol, their partition into the DPPC-rich
$L_o$ phase depends on the associate exchange parameter. This is the
reason why the partition behavior of the proteins changes rapidly when
$\varepsilon_{25}\simeq\varepsilon_{22}=1.3\varepsilon$, as can be
seen in fig.~\ref{fig:2phase} summarizing our results from simulations
of mixtures in the two-phase region. Model proteins with low affinity
to saturated lipids remain in the DPPC-poor $L_d$ phase and,
conversely, when $\varepsilon_{25}>\varepsilon_{22}$ they accumulate
in the DPPC-rich $L_o$ phase.

At low compositions of saturated DPPC lipids, the ternary mixture is
one phase which is predominantly liquid-disordered, with local density
fluctuations appearing as liquid-ordered domains of nanometric
scale. These dynamic domains are 1-1.5 order of magnitude smaller than
lipid rafts. The presence of proteins in biological membranes has been
mentioned as one of the factors contributing to the growth and
meta-stability of lipid rafts. Our model simulations reveal that the
presence of even a small density of small non-interacting peptides
can, indeed, lead to dramatic changes in these properties of the
liquid-ordered domains formed in ternary mixtures. We find (see
fig.~\ref{Fig:12dppc}) that with the increase in the affinity
parameter $\varepsilon_{25}$, the liquid ordered domains grow in size
by recruiting proteins and saturated lipids. These large
liquid-ordered domains become increasingly metastable and begin to
develop gel-like clusters which are blends of lipids and ordered
saturated lipid chains. Eventually, when the affinity between the
proteins and saturated lipids exceeds the affinity of the saturated
lipids to each other, the liquid-ordered domains merge into a single
distinct stable phase containing most of the proteins. Depending on
the relative proportions of proteins and saturated lipids, the phase
separated from the $L_d$ environment may be either $L_o$ or $S_o$ (see
fig.~\ref{fig:score12dppc}). The biologically most relevant result may
be the behavior of the 18DPPC mixtures when
$\varepsilon_{25}\lesssim\varepsilon_{22}$ [see, e.g.,
  fig.~\ref{fig:score18dppc}, snapshots (b)-(c)]. There, near the
critical point of the ternary mixture, the presence of low density of
proteins, can lead to the formation of metastable dynamic
liquid-ordered domains whose size is comparable to the size of lipid
rafts (see Supplementary Material, SI movie 18DPPC\_1.3.mp4).

We conclude by reminding the reader that cell membranes are larger and
far more complex than the model mixtures studied herein. They include
larger proteins of different kinds, and their dynamic behavior depends
on many more factor, including non-equilibrium contributions arising,
e.g., from their interactions with the cell cytoskeleton. Different
regions of these membrane may be effectively subject to different
local compositions and interactions. Our study shows that even a
simple mixture model with a small number of interaction parameters,
may exhibit rich equilibrium phase behavior with characteristics
resembling key features of lipid rafts.

\section*{Supplementary Material}

See supplementary material for movies showing the dynamics of
different mixtures.

\begin{acknowledgements}
This work was supported by the Israel Science Foundation (ISF), grant
No.~1258/22.
\end{acknowledgements}
\section*{Data Availability Statement}
The data validating the findings of our simulation study are available
upon request to the corresponding author.
\section*{Conflict of Interest}
The authors declare no conflict of interest.
\section*{Author Contributions}
\textbf{SB:} Software; Data curation; Analysis;
Writing. \textbf{OF:} Conceptualization; supervision; Funding
Acquisition; Writing.
%\clearpage
\nocite{*}
%\bibliography{aipsamp}% Produces the bibliography via BibTeX.
\bibliography{library2}

\begin{thebibliography}{102}%
\makeatletter
\providecommand \@ifxundefined [1]{%
 \@ifx{#1\undefined}
}%
\providecommand \@ifnum [1]{%
 \ifnum #1\expandafter \@firstoftwo
 \else \expandafter \@secondoftwo
 \fi
}%
\providecommand \@ifx [1]{%
 \ifx #1\expandafter \@firstoftwo
 \else \expandafter \@secondoftwo
 \fi
}%
\providecommand \natexlab [1]{#1}%
\providecommand \enquote  [1]{``#1''}%
\providecommand \bibnamefont  [1]{#1}%
\providecommand \bibfnamefont [1]{#1}%
\providecommand \citenamefont [1]{#1}%
\providecommand \href@noop [0]{\@secondoftwo}%
\providecommand \href [0]{\begingroup \@sanitize@url \@href}%
\providecommand \@href[1]{\@@startlink{#1}\@@href}%
\providecommand \@@href[1]{\endgroup#1\@@endlink}%
\providecommand \@sanitize@url [0]{\catcode `\\12\catcode `\$12\catcode `\&12\catcode `\#12\catcode `\^12\catcode `\_12\catcode `\%12\relax}%
\providecommand \@@startlink[1]{}%
\providecommand \@@endlink[0]{}%
\providecommand \url  [0]{\begingroup\@sanitize@url \@url }%
\providecommand \@url [1]{\endgroup\@href {#1}{\urlprefix }}%
\providecommand \urlprefix  [0]{URL }%
\providecommand \Eprint [0]{\href }%
\providecommand \doibase [0]{http://dx.doi.org/}%
\providecommand \selectlanguage [0]{\@gobble}%
\providecommand \bibinfo  [0]{\@secondoftwo}%
\providecommand \bibfield  [0]{\@secondoftwo}%
\providecommand \translation [1]{[#1]}%
\providecommand \BibitemOpen [0]{}%
\providecommand \bibitemStop [0]{}%
\providecommand \bibitemNoStop [0]{.\EOS\space}%
\providecommand \EOS [0]{\spacefactor3000\relax}%
\providecommand \BibitemShut  [1]{\csname bibitem#1\endcsname}%
\let\auto@bib@innerbib\@empty
%</preamble>
\bibitem [{\citenamefont {Alberts}(2017)}]{Alberts2017}%
  \BibitemOpen
  \bibfield  {author} {\bibinfo {author} {\bibfnamefont {B.}~\bibnamefont {Alberts}},\ }\href@noop {} {\emph {\bibinfo {title} {Molecular biology of the cell}}}\ (\bibinfo  {publisher} {WW Norton \& Company},\ \bibinfo {year} {2017})\BibitemShut {NoStop}%
\bibitem [{\citenamefont {Cheng}\ and\ \citenamefont {Smith}(2019)}]{cheng2019biological}%
  \BibitemOpen
  \bibfield  {author} {\bibinfo {author} {\bibfnamefont {X.}~\bibnamefont {Cheng}}\ and\ \bibinfo {author} {\bibfnamefont {J.~C.}\ \bibnamefont {Smith}},\ }\href@noop {} {\bibfield  {journal} {\bibinfo  {journal} {Chem. Rev.}\ }\textbf {\bibinfo {volume} {119}},\ \bibinfo {pages} {5849} (\bibinfo {year} {2019})}\BibitemShut {NoStop}%
\bibitem [{\citenamefont {Shevchenko}\ and\ \citenamefont {Simons}(2010)}]{shevchenko2010lipidomics}%
  \BibitemOpen
  \bibfield  {author} {\bibinfo {author} {\bibfnamefont {A.}~\bibnamefont {Shevchenko}}\ and\ \bibinfo {author} {\bibfnamefont {K.}~\bibnamefont {Simons}},\ }\href@noop {} {\bibfield  {journal} {\bibinfo  {journal} {Nat. Rev. Mol. Cell Biol.}\ }\textbf {\bibinfo {volume} {11}},\ \bibinfo {pages} {593} (\bibinfo {year} {2010})}\BibitemShut {NoStop}%
\bibitem [{\citenamefont {xiao Shen}\ \emph {et~al.}(2014)\citenamefont {xiao Shen}, \citenamefont {Saboe}, \citenamefont {Sines}, \citenamefont {Erbakan},\ and\ \citenamefont {Kumar}}]{SHEN2014359}%
  \BibitemOpen
  \bibfield  {author} {\bibinfo {author} {\bibfnamefont {Y.}~\bibnamefont {xiao Shen}}, \bibinfo {author} {\bibfnamefont {P.~O.}\ \bibnamefont {Saboe}}, \bibinfo {author} {\bibfnamefont {I.~T.}\ \bibnamefont {Sines}}, \bibinfo {author} {\bibfnamefont {M.}~\bibnamefont {Erbakan}}, \ and\ \bibinfo {author} {\bibfnamefont {M.}~\bibnamefont {Kumar}},\ }\href {\doibase https://doi.org/10.1016/j.memsci.2013.12.019} {\bibfield  {journal} {\bibinfo  {journal} {J. Membr. Sci.}\ }\textbf {\bibinfo {volume} {454}},\ \bibinfo {pages} {359} (\bibinfo {year} {2014})}\BibitemShut {NoStop}%
\bibitem [{\citenamefont {Blanco}\ and\ \citenamefont {Blanco}(2017)}]{BLANCO2017215}%
  \BibitemOpen
  \bibfield  {author} {\bibinfo {author} {\bibfnamefont {A.}~\bibnamefont {Blanco}}\ and\ \bibinfo {author} {\bibfnamefont {G.}~\bibnamefont {Blanco}},\ }in\ \href {\doibase https://doi.org/10.1016/B978-0-12-803550-4.00011-2} {\emph {\bibinfo {booktitle} {Med. Biochem.}}},\ \bibinfo {editor} {edited by\ \bibinfo {editor} {\bibfnamefont {A.}~\bibnamefont {Blanco}}\ and\ \bibinfo {editor} {\bibfnamefont {G.}~\bibnamefont {Blanco}}}\ (\bibinfo  {publisher} {Academic Press},\ \bibinfo {year} {2017})\ pp.\ \bibinfo {pages} {215--250}\BibitemShut {NoStop}%
\bibitem [{\citenamefont {Guigas}\ and\ \citenamefont {Weiss}(2016)}]{guigas2016effects}%
  \BibitemOpen
  \bibfield  {author} {\bibinfo {author} {\bibfnamefont {G.}~\bibnamefont {Guigas}}\ and\ \bibinfo {author} {\bibfnamefont {M.}~\bibnamefont {Weiss}},\ }\href@noop {} {\bibfield  {journal} {\bibinfo  {journal} {Biochim. Biophys. Acta, Biomembr.}\ }\textbf {\bibinfo {volume} {1858}},\ \bibinfo {pages} {2441} (\bibinfo {year} {2016})}\BibitemShut {NoStop}%
\bibitem [{\citenamefont {Konings}, \citenamefont {Kaback},\ and\ \citenamefont {Lolkema}(1996)}]{konings1996transport}%
  \BibitemOpen
  \bibfield  {author} {\bibinfo {author} {\bibfnamefont {W.~N.}\ \bibnamefont {Konings}}, \bibinfo {author} {\bibfnamefont {H.~R.}\ \bibnamefont {Kaback}}, \ and\ \bibinfo {author} {\bibfnamefont {J.~S.}\ \bibnamefont {Lolkema}},\ }\href@noop {} {\emph {\bibinfo {title} {Transport processes in eukaryotic and prokaryotic organisms}}}\ (\bibinfo  {publisher} {Elsevier},\ \bibinfo {year} {1996})\BibitemShut {NoStop}%
\bibitem [{\citenamefont {Jelokhani-Niaraki}(2022)}]{jelokhani2022membrane}%
  \BibitemOpen
  \bibfield  {author} {\bibinfo {author} {\bibfnamefont {M.}~\bibnamefont {Jelokhani-Niaraki}},\ }\href@noop {} {\enquote {\bibinfo {title} {Membrane proteins: structure, function and motion},}\ } (\bibinfo {year} {2022})\BibitemShut {NoStop}%
\bibitem [{\citenamefont {Whitford}(2013)}]{whitford2013proteins}%
  \BibitemOpen
  \bibfield  {author} {\bibinfo {author} {\bibfnamefont {D.}~\bibnamefont {Whitford}},\ }\href@noop {} {\emph {\bibinfo {title} {Proteins: structure and function}}}\ (\bibinfo  {publisher} {John Wiley \& Sons},\ \bibinfo {year} {2013})\BibitemShut {NoStop}%
\bibitem [{\citenamefont {Sj}(1972)}]{sj1972fluid}%
  \BibitemOpen
  \bibfield  {author} {\bibinfo {author} {\bibfnamefont {S.}~\bibnamefont {Sj}},\ }\href@noop {} {\bibfield  {journal} {\bibinfo  {journal} {Science}\ }\textbf {\bibinfo {volume} {175}},\ \bibinfo {pages} {720} (\bibinfo {year} {1972})}\BibitemShut {NoStop}%
\bibitem [{\citenamefont {Ahmed}, \citenamefont {Brown},\ and\ \citenamefont {London}(1997)}]{ahmed1997origin}%
  \BibitemOpen
  \bibfield  {author} {\bibinfo {author} {\bibfnamefont {S.~N.}\ \bibnamefont {Ahmed}}, \bibinfo {author} {\bibfnamefont {D.~A.}\ \bibnamefont {Brown}}, \ and\ \bibinfo {author} {\bibfnamefont {E.}~\bibnamefont {London}},\ }\href@noop {} {\bibfield  {journal} {\bibinfo  {journal} {Biochemistry}\ }\textbf {\bibinfo {volume} {36}},\ \bibinfo {pages} {10944} (\bibinfo {year} {1997})}\BibitemShut {NoStop}%
\bibitem [{\citenamefont {Brown}\ and\ \citenamefont {Rose}(1992)}]{brown1992sorting}%
  \BibitemOpen
  \bibfield  {author} {\bibinfo {author} {\bibfnamefont {D.~A.}\ \bibnamefont {Brown}}\ and\ \bibinfo {author} {\bibfnamefont {J.~K.}\ \bibnamefont {Rose}},\ }\href@noop {} {\bibfield  {journal} {\bibinfo  {journal} {Cell}\ }\textbf {\bibinfo {volume} {68}},\ \bibinfo {pages} {533} (\bibinfo {year} {1992})}\BibitemShut {NoStop}%
\bibitem [{\citenamefont {Pralle}\ \emph {et~al.}(2000)\citenamefont {Pralle}, \citenamefont {Keller}, \citenamefont {Florin}, \citenamefont {Simons},\ and\ \citenamefont {H{\"o}rber}}]{pralle2000sphingolipid}%
  \BibitemOpen
  \bibfield  {author} {\bibinfo {author} {\bibfnamefont {A.}~\bibnamefont {Pralle}}, \bibinfo {author} {\bibfnamefont {P.}~\bibnamefont {Keller}}, \bibinfo {author} {\bibfnamefont {E.-L.}\ \bibnamefont {Florin}}, \bibinfo {author} {\bibfnamefont {K.}~\bibnamefont {Simons}}, \ and\ \bibinfo {author} {\bibfnamefont {J.~H.}\ \bibnamefont {H{\"o}rber}},\ }\href@noop {} {\bibfield  {journal} {\bibinfo  {journal} {J. Cell Biol.}\ }\textbf {\bibinfo {volume} {148}},\ \bibinfo {pages} {997} (\bibinfo {year} {2000})}\BibitemShut {NoStop}%
\bibitem [{\citenamefont {Ackerman}\ and\ \citenamefont {Feigenson}(2015)}]{ackerman2015lipid}%
  \BibitemOpen
  \bibfield  {author} {\bibinfo {author} {\bibfnamefont {D.~G.}\ \bibnamefont {Ackerman}}\ and\ \bibinfo {author} {\bibfnamefont {G.~W.}\ \bibnamefont {Feigenson}},\ }\href@noop {} {\bibfield  {journal} {\bibinfo  {journal} {Essays Biochem.}\ }\textbf {\bibinfo {volume} {57}},\ \bibinfo {pages} {33} (\bibinfo {year} {2015})}\BibitemShut {NoStop}%
\bibitem [{\citenamefont {Huang}\ and\ \citenamefont {Feigenson}(1999)}]{huang1999microscopic}%
  \BibitemOpen
  \bibfield  {author} {\bibinfo {author} {\bibfnamefont {J.}~\bibnamefont {Huang}}\ and\ \bibinfo {author} {\bibfnamefont {G.~W.}\ \bibnamefont {Feigenson}},\ }\href@noop {} {\bibfield  {journal} {\bibinfo  {journal} {Biophys. J.}\ }\textbf {\bibinfo {volume} {76}},\ \bibinfo {pages} {2142} (\bibinfo {year} {1999})}\BibitemShut {NoStop}%
\bibitem [{\citenamefont {Schuck}\ \emph {et~al.}(2003)\citenamefont {Schuck}, \citenamefont {Honsho}, \citenamefont {Ekroos}, \citenamefont {Shevchenko},\ and\ \citenamefont {Simons}}]{schuck2003resistance}%
  \BibitemOpen
  \bibfield  {author} {\bibinfo {author} {\bibfnamefont {S.}~\bibnamefont {Schuck}}, \bibinfo {author} {\bibfnamefont {M.}~\bibnamefont {Honsho}}, \bibinfo {author} {\bibfnamefont {K.}~\bibnamefont {Ekroos}}, \bibinfo {author} {\bibfnamefont {A.}~\bibnamefont {Shevchenko}}, \ and\ \bibinfo {author} {\bibfnamefont {K.}~\bibnamefont {Simons}},\ }\href@noop {} {\bibfield  {journal} {\bibinfo  {journal} {Proc. Natl. Acad. Sci.}\ }\textbf {\bibinfo {volume} {100}},\ \bibinfo {pages} {5795} (\bibinfo {year} {2003})}\BibitemShut {NoStop}%
\bibitem [{\citenamefont {Pike}(2006)}]{pike2006rafts}%
  \BibitemOpen
  \bibfield  {author} {\bibinfo {author} {\bibfnamefont {L.~J.}\ \bibnamefont {Pike}},\ }\href@noop {} {\bibfield  {journal} {\bibinfo  {journal} {J. Lipid Res.}\ }\textbf {\bibinfo {volume} {47}},\ \bibinfo {pages} {1597} (\bibinfo {year} {2006})}\BibitemShut {NoStop}%
\bibitem [{\citenamefont {Mouritsen}(2010)}]{mouritsen2010liquid}%
  \BibitemOpen
  \bibfield  {author} {\bibinfo {author} {\bibfnamefont {O.~G.}\ \bibnamefont {Mouritsen}},\ }\href@noop {} {\bibfield  {journal} {\bibinfo  {journal} {Biochim. Biophys. Acta, Biomembr.}\ }\textbf {\bibinfo {volume} {1798}},\ \bibinfo {pages} {1286} (\bibinfo {year} {2010})}\BibitemShut {NoStop}%
\bibitem [{\citenamefont {Filippov}, \citenamefont {Or{\"a}dd},\ and\ \citenamefont {Lindblom}(2004)}]{filippov2004lipid}%
  \BibitemOpen
  \bibfield  {author} {\bibinfo {author} {\bibfnamefont {A.}~\bibnamefont {Filippov}}, \bibinfo {author} {\bibfnamefont {G.}~\bibnamefont {Or{\"a}dd}}, \ and\ \bibinfo {author} {\bibfnamefont {G.}~\bibnamefont {Lindblom}},\ }\href@noop {} {\bibfield  {journal} {\bibinfo  {journal} {Biophys. J.}\ }\textbf {\bibinfo {volume} {86}},\ \bibinfo {pages} {891} (\bibinfo {year} {2004})}\BibitemShut {NoStop}%
\bibitem [{\citenamefont {Holl}(2008)}]{holl2008cell}%
  \BibitemOpen
  \bibfield  {author} {\bibinfo {author} {\bibfnamefont {M.~M.~B.}\ \bibnamefont {Holl}},\ }\enquote {\bibinfo {title} {Cell plasma membranes and phase transitions},}\ in\ \href {\doibase 10.1007/978-1-4020-8651-9_12} {\emph {\bibinfo {booktitle} {Phase Transitions in Cell Biology}}},\ \bibinfo {editor} {edited by\ \bibinfo {editor} {\bibfnamefont {G.~H.}\ \bibnamefont {Pollack}}\ and\ \bibinfo {editor} {\bibfnamefont {W.-C.}\ \bibnamefont {Chin}}}\ (\bibinfo  {publisher} {Springer Netherlands},\ \bibinfo {address} {Dordrecht},\ \bibinfo {year} {2008})\ pp.\ \bibinfo {pages} {171--181}\BibitemShut {NoStop}%
\bibitem [{\citenamefont {Marsh}(2013)}]{marsh2013handbook}%
  \BibitemOpen
  \bibfield  {author} {\bibinfo {author} {\bibfnamefont {D.}~\bibnamefont {Marsh}},\ }\href@noop {} {\emph {\bibinfo {title} {Handbook of lipid bilayers}}}\ (\bibinfo  {publisher} {CRC press},\ \bibinfo {year} {2013})\BibitemShut {NoStop}%
\bibitem [{\citenamefont {Feigenson}(2009)}]{feigenson2009phase}%
  \BibitemOpen
  \bibfield  {author} {\bibinfo {author} {\bibfnamefont {G.~W.}\ \bibnamefont {Feigenson}},\ }\href@noop {} {\bibfield  {journal} {\bibinfo  {journal} {Biochim. Biophys. Acta, Biomembr.}\ }\textbf {\bibinfo {volume} {1788}},\ \bibinfo {pages} {47} (\bibinfo {year} {2009})}\BibitemShut {NoStop}%
\bibitem [{\citenamefont {Komura}\ and\ \citenamefont {Andelman}(2014)}]{komura2014physical}%
  \BibitemOpen
  \bibfield  {author} {\bibinfo {author} {\bibfnamefont {S.}~\bibnamefont {Komura}}\ and\ \bibinfo {author} {\bibfnamefont {D.}~\bibnamefont {Andelman}},\ }\href@noop {} {\bibfield  {journal} {\bibinfo  {journal} {Adv. Colloid Interface Sci.}\ }\textbf {\bibinfo {volume} {208}},\ \bibinfo {pages} {34} (\bibinfo {year} {2014})}\BibitemShut {NoStop}%
\bibitem [{\citenamefont {Veatch}\ \emph {et~al.}(2007)\citenamefont {Veatch}, \citenamefont {Soubias}, \citenamefont {Keller},\ and\ \citenamefont {Gawrisch}}]{veatch2007critical}%
  \BibitemOpen
  \bibfield  {author} {\bibinfo {author} {\bibfnamefont {S.~L.}\ \bibnamefont {Veatch}}, \bibinfo {author} {\bibfnamefont {O.}~\bibnamefont {Soubias}}, \bibinfo {author} {\bibfnamefont {S.~L.}\ \bibnamefont {Keller}}, \ and\ \bibinfo {author} {\bibfnamefont {K.}~\bibnamefont {Gawrisch}},\ }\href@noop {} {\bibfield  {journal} {\bibinfo  {journal} {Proc. Natl. Acad. Sci.}\ }\textbf {\bibinfo {volume} {104}},\ \bibinfo {pages} {17650} (\bibinfo {year} {2007})}\BibitemShut {NoStop}%
\bibitem [{\citenamefont {Hirst}, \citenamefont {Uppamoochikkal},\ and\ \citenamefont {Lor}(2011)}]{hirst2011phase}%
  \BibitemOpen
  \bibfield  {author} {\bibinfo {author} {\bibfnamefont {L.~S.}\ \bibnamefont {Hirst}}, \bibinfo {author} {\bibfnamefont {P.}~\bibnamefont {Uppamoochikkal}}, \ and\ \bibinfo {author} {\bibfnamefont {C.}~\bibnamefont {Lor}},\ }\href@noop {} {\bibfield  {journal} {\bibinfo  {journal} {Liq. Cryst.}\ }\textbf {\bibinfo {volume} {38}},\ \bibinfo {pages} {1735} (\bibinfo {year} {2011})}\BibitemShut {NoStop}%
\bibitem [{\citenamefont {St{\"o}ckl}\ and\ \citenamefont {Herrmann}(2010)}]{stockl2010detection}%
  \BibitemOpen
  \bibfield  {author} {\bibinfo {author} {\bibfnamefont {M.}~\bibnamefont {St{\"o}ckl}}\ and\ \bibinfo {author} {\bibfnamefont {A.}~\bibnamefont {Herrmann}},\ }\href@noop {} {\bibfield  {journal} {\bibinfo  {journal} {Biochim. Biophys. Acta, Biomembr.}\ }\textbf {\bibinfo {volume} {1798}},\ \bibinfo {pages} {1444} (\bibinfo {year} {2010})}\BibitemShut {NoStop}%
\bibitem [{\citenamefont {Bagatolli}\ and\ \citenamefont {Gratton}(2001)}]{bagatolli2001direct}%
  \BibitemOpen
  \bibfield  {author} {\bibinfo {author} {\bibfnamefont {L.~A.}\ \bibnamefont {Bagatolli}}\ and\ \bibinfo {author} {\bibfnamefont {E.}~\bibnamefont {Gratton}},\ }\href@noop {} {\bibfield  {journal} {\bibinfo  {journal} {J. Fluoresc.}\ }\textbf {\bibinfo {volume} {11}},\ \bibinfo {pages} {141} (\bibinfo {year} {2001})}\BibitemShut {NoStop}%
\bibitem [{\citenamefont {Bagatolli}\ and\ \citenamefont {Gratton}(2000)}]{bagatolli2000correlation}%
  \BibitemOpen
  \bibfield  {author} {\bibinfo {author} {\bibfnamefont {L.~A.}\ \bibnamefont {Bagatolli}}\ and\ \bibinfo {author} {\bibfnamefont {E.}~\bibnamefont {Gratton}},\ }\href@noop {} {\bibfield  {journal} {\bibinfo  {journal} {Biophys. J.}\ }\textbf {\bibinfo {volume} {79}},\ \bibinfo {pages} {434} (\bibinfo {year} {2000})}\BibitemShut {NoStop}%
\bibitem [{\citenamefont {Farago}(2021)}]{farago2021beginner}%
  \BibitemOpen
  \bibfield  {author} {\bibinfo {author} {\bibfnamefont {O.}~\bibnamefont {Farago}},\ }in\ \href@noop {} {\emph {\bibinfo {booktitle} {Modeling Biomaterials}}}\ (\bibinfo  {publisher} {Springer},\ \bibinfo {year} {2021})\ pp.\ \bibinfo {pages} {1--41}\BibitemShut {NoStop}%
\bibitem [{\citenamefont {Koukalov{\'a}}\ \emph {et~al.}(2017)\citenamefont {Koukalov{\'a}}, \citenamefont {Amaro}, \citenamefont {Aydogan}, \citenamefont {Gr{\"o}bner}, \citenamefont {Williamson}, \citenamefont {Mikhalyov}, \citenamefont {Hof},\ and\ \citenamefont {{\v{S}}achl}}]{koukalova2017lipid}%
  \BibitemOpen
  \bibfield  {author} {\bibinfo {author} {\bibfnamefont {A.}~\bibnamefont {Koukalov{\'a}}}, \bibinfo {author} {\bibfnamefont {M.}~\bibnamefont {Amaro}}, \bibinfo {author} {\bibfnamefont {G.}~\bibnamefont {Aydogan}}, \bibinfo {author} {\bibfnamefont {G.}~\bibnamefont {Gr{\"o}bner}}, \bibinfo {author} {\bibfnamefont {P.~T.}\ \bibnamefont {Williamson}}, \bibinfo {author} {\bibfnamefont {I.}~\bibnamefont {Mikhalyov}}, \bibinfo {author} {\bibfnamefont {M.}~\bibnamefont {Hof}}, \ and\ \bibinfo {author} {\bibfnamefont {R.}~\bibnamefont {{\v{S}}achl}},\ }\href@noop {} {\bibfield  {journal} {\bibinfo  {journal} {Sci. Rep.}\ }\textbf {\bibinfo {volume} {7}},\ \bibinfo {pages} {5460} (\bibinfo {year} {2017})}\BibitemShut {NoStop}%
\bibitem [{\citenamefont {de~Almeida}\ \emph {et~al.}(2005)\citenamefont {de~Almeida}, \citenamefont {Loura}, \citenamefont {Fedorov},\ and\ \citenamefont {Prieto}}]{de2005lipid}%
  \BibitemOpen
  \bibfield  {author} {\bibinfo {author} {\bibfnamefont {R.~F.}\ \bibnamefont {de~Almeida}}, \bibinfo {author} {\bibfnamefont {L.~M.}\ \bibnamefont {Loura}}, \bibinfo {author} {\bibfnamefont {A.}~\bibnamefont {Fedorov}}, \ and\ \bibinfo {author} {\bibfnamefont {M.}~\bibnamefont {Prieto}},\ }\href@noop {} {\bibfield  {journal} {\bibinfo  {journal} {J. Mol. Biol.}\ }\textbf {\bibinfo {volume} {346}},\ \bibinfo {pages} {1109} (\bibinfo {year} {2005})}\BibitemShut {NoStop}%
\bibitem [{\citenamefont {{\v{S}}achl}\ \emph {et~al.}(2011)\citenamefont {{\v{S}}achl}, \citenamefont {Humpol{\'\i}{\v{c}}kov{\'a}}, \citenamefont {{\v{S}}tefl}, \citenamefont {Johansson},\ and\ \citenamefont {Hof}}]{vsachl2011limitations}%
  \BibitemOpen
  \bibfield  {author} {\bibinfo {author} {\bibfnamefont {R.}~\bibnamefont {{\v{S}}achl}}, \bibinfo {author} {\bibfnamefont {J.}~\bibnamefont {Humpol{\'\i}{\v{c}}kov{\'a}}}, \bibinfo {author} {\bibfnamefont {M.}~\bibnamefont {{\v{S}}tefl}}, \bibinfo {author} {\bibfnamefont {L.~B.-{\AA}.}\ \bibnamefont {Johansson}}, \ and\ \bibinfo {author} {\bibfnamefont {M.}~\bibnamefont {Hof}},\ }\href@noop {} {\bibfield  {journal} {\bibinfo  {journal} {Biophys. J.}\ }\textbf {\bibinfo {volume} {101}},\ \bibinfo {pages} {L60} (\bibinfo {year} {2011})}\BibitemShut {NoStop}%
\bibitem [{\citenamefont {De~Wit}\ \emph {et~al.}(2015)\citenamefont {De~Wit}, \citenamefont {Danial}, \citenamefont {Kukura},\ and\ \citenamefont {Wallace}}]{de2015dynamic}%
  \BibitemOpen
  \bibfield  {author} {\bibinfo {author} {\bibfnamefont {G.}~\bibnamefont {De~Wit}}, \bibinfo {author} {\bibfnamefont {J.~S.}\ \bibnamefont {Danial}}, \bibinfo {author} {\bibfnamefont {P.}~\bibnamefont {Kukura}}, \ and\ \bibinfo {author} {\bibfnamefont {M.~I.}\ \bibnamefont {Wallace}},\ }\href@noop {} {\bibfield  {journal} {\bibinfo  {journal} {Proc. Natl. Acad. Sci.}\ }\textbf {\bibinfo {volume} {112}},\ \bibinfo {pages} {12299} (\bibinfo {year} {2015})}\BibitemShut {NoStop}%
\bibitem [{\citenamefont {Giocondi}\ \emph {et~al.}(2001)\citenamefont {Giocondi}, \citenamefont {Vi{\'e}}, \citenamefont {Lesniewska}, \citenamefont {Milhiet}, \citenamefont {Zinke-Allmang},\ and\ \citenamefont {Le~Grimellec}}]{giocondi2001phase}%
  \BibitemOpen
  \bibfield  {author} {\bibinfo {author} {\bibfnamefont {M.-C.}\ \bibnamefont {Giocondi}}, \bibinfo {author} {\bibfnamefont {V.}~\bibnamefont {Vi{\'e}}}, \bibinfo {author} {\bibfnamefont {E.}~\bibnamefont {Lesniewska}}, \bibinfo {author} {\bibfnamefont {P.-E.}\ \bibnamefont {Milhiet}}, \bibinfo {author} {\bibfnamefont {M.}~\bibnamefont {Zinke-Allmang}}, \ and\ \bibinfo {author} {\bibfnamefont {C.}~\bibnamefont {Le~Grimellec}},\ }\href@noop {} {\bibfield  {journal} {\bibinfo  {journal} {Langmuir}\ }\textbf {\bibinfo {volume} {17}},\ \bibinfo {pages} {1653} (\bibinfo {year} {2001})}\BibitemShut {NoStop}%
\bibitem [{\citenamefont {Tokumasu}\ \emph {et~al.}(2003)\citenamefont {Tokumasu}, \citenamefont {Jin}, \citenamefont {Feigenson},\ and\ \citenamefont {Dvorak}}]{tokumasu2003nanoscopic}%
  \BibitemOpen
  \bibfield  {author} {\bibinfo {author} {\bibfnamefont {F.}~\bibnamefont {Tokumasu}}, \bibinfo {author} {\bibfnamefont {A.~J.}\ \bibnamefont {Jin}}, \bibinfo {author} {\bibfnamefont {G.~W.}\ \bibnamefont {Feigenson}}, \ and\ \bibinfo {author} {\bibfnamefont {J.~A.}\ \bibnamefont {Dvorak}},\ }\href@noop {} {\bibfield  {journal} {\bibinfo  {journal} {Biophys. J.}\ }\textbf {\bibinfo {volume} {84}},\ \bibinfo {pages} {2609} (\bibinfo {year} {2003})}\BibitemShut {NoStop}%
\bibitem [{\citenamefont {Choucair}\ \emph {et~al.}(2007)\citenamefont {Choucair}, \citenamefont {Chakrapani}, \citenamefont {Chakravarthy}, \citenamefont {Katsaras},\ and\ \citenamefont {Johnston}}]{choucair2007preferential}%
  \BibitemOpen
  \bibfield  {author} {\bibinfo {author} {\bibfnamefont {A.}~\bibnamefont {Choucair}}, \bibinfo {author} {\bibfnamefont {M.}~\bibnamefont {Chakrapani}}, \bibinfo {author} {\bibfnamefont {B.}~\bibnamefont {Chakravarthy}}, \bibinfo {author} {\bibfnamefont {J.}~\bibnamefont {Katsaras}}, \ and\ \bibinfo {author} {\bibfnamefont {L.}~\bibnamefont {Johnston}},\ }\href@noop {} {\bibfield  {journal} {\bibinfo  {journal} {Biochim. Biophys. Acta, Biomembr.}\ }\textbf {\bibinfo {volume} {1768}},\ \bibinfo {pages} {146} (\bibinfo {year} {2007})}\BibitemShut {NoStop}%
\bibitem [{\citenamefont {Veatch}\ \emph {et~al.}(2004)\citenamefont {Veatch}, \citenamefont {Polozov}, \citenamefont {Gawrisch},\ and\ \citenamefont {Keller}}]{veatch2004liquid}%
  \BibitemOpen
  \bibfield  {author} {\bibinfo {author} {\bibfnamefont {S.~L.}\ \bibnamefont {Veatch}}, \bibinfo {author} {\bibfnamefont {I.}~\bibnamefont {Polozov}}, \bibinfo {author} {\bibfnamefont {K.}~\bibnamefont {Gawrisch}}, \ and\ \bibinfo {author} {\bibfnamefont {S.~L.}\ \bibnamefont {Keller}},\ }\href@noop {} {\bibfield  {journal} {\bibinfo  {journal} {Biophys. J.}\ }\textbf {\bibinfo {volume} {86}},\ \bibinfo {pages} {2910} (\bibinfo {year} {2004})}\BibitemShut {NoStop}%
\bibitem [{\citenamefont {Marrink}\ \emph {et~al.}(2019)\citenamefont {Marrink}, \citenamefont {Corradi}, \citenamefont {Souza}, \citenamefont {Ingolfsson}, \citenamefont {Tieleman},\ and\ \citenamefont {Sansom}}]{marrink2019computational}%
  \BibitemOpen
  \bibfield  {author} {\bibinfo {author} {\bibfnamefont {S.~J.}\ \bibnamefont {Marrink}}, \bibinfo {author} {\bibfnamefont {V.}~\bibnamefont {Corradi}}, \bibinfo {author} {\bibfnamefont {P.~C.}\ \bibnamefont {Souza}}, \bibinfo {author} {\bibfnamefont {H.~I.}\ \bibnamefont {Ingolfsson}}, \bibinfo {author} {\bibfnamefont {D.~P.}\ \bibnamefont {Tieleman}}, \ and\ \bibinfo {author} {\bibfnamefont {M.~S.}\ \bibnamefont {Sansom}},\ }\href@noop {} {\bibfield  {journal} {\bibinfo  {journal} {Chem. Rev.}\ }\textbf {\bibinfo {volume} {119}},\ \bibinfo {pages} {6184} (\bibinfo {year} {2019})}\BibitemShut {NoStop}%
\bibitem [{\citenamefont {Bennett}\ and\ \citenamefont {Tieleman}(2013{\natexlab{a}})}]{BENNETT20131765}%
  \BibitemOpen
  \bibfield  {author} {\bibinfo {author} {\bibfnamefont {W.~D.}\ \bibnamefont {Bennett}}\ and\ \bibinfo {author} {\bibfnamefont {D.~P.}\ \bibnamefont {Tieleman}},\ }\href {\doibase https://doi.org/10.1016/j.bbamem.2013.03.004} {\bibfield  {journal} {\bibinfo  {journal} {Biochim. Biophys. Acta, Biomembr}\ }\textbf {\bibinfo {volume} {1828}},\ \bibinfo {pages} {1765} (\bibinfo {year} {2013}{\natexlab{a}})}\BibitemShut {NoStop}%
\bibitem [{\citenamefont {Bennett}\ and\ \citenamefont {Tieleman}(2013{\natexlab{b}})}]{bennett2013computer}%
  \BibitemOpen
  \bibfield  {author} {\bibinfo {author} {\bibfnamefont {W.~D.}\ \bibnamefont {Bennett}}\ and\ \bibinfo {author} {\bibfnamefont {D.~P.}\ \bibnamefont {Tieleman}},\ }\href@noop {} {\bibfield  {journal} {\bibinfo  {journal} {Biochim. Biophys. Acta, Biomembr.}\ }\textbf {\bibinfo {volume} {1828}},\ \bibinfo {pages} {1765} (\bibinfo {year} {2013}{\natexlab{b}})}\BibitemShut {NoStop}%
\bibitem [{\citenamefont {Javanainen}, \citenamefont {Martinez-Seara},\ and\ \citenamefont {Vattulainen}(2017)}]{javanainen2017nanoscale}%
  \BibitemOpen
  \bibfield  {author} {\bibinfo {author} {\bibfnamefont {M.}~\bibnamefont {Javanainen}}, \bibinfo {author} {\bibfnamefont {H.}~\bibnamefont {Martinez-Seara}}, \ and\ \bibinfo {author} {\bibfnamefont {I.}~\bibnamefont {Vattulainen}},\ }\href@noop {} {\bibfield  {journal} {\bibinfo  {journal} {Sci. Rep.}\ }\textbf {\bibinfo {volume} {7}},\ \bibinfo {pages} {1143} (\bibinfo {year} {2017})}\BibitemShut {NoStop}%
\bibitem [{\citenamefont {Sodt}\ \emph {et~al.}(2014)\citenamefont {Sodt}, \citenamefont {Sandar}, \citenamefont {Gawrisch}, \citenamefont {Pastor},\ and\ \citenamefont {Lyman}}]{sodt2014molecular}%
  \BibitemOpen
  \bibfield  {author} {\bibinfo {author} {\bibfnamefont {A.~J.}\ \bibnamefont {Sodt}}, \bibinfo {author} {\bibfnamefont {M.~L.}\ \bibnamefont {Sandar}}, \bibinfo {author} {\bibfnamefont {K.}~\bibnamefont {Gawrisch}}, \bibinfo {author} {\bibfnamefont {R.~W.}\ \bibnamefont {Pastor}}, \ and\ \bibinfo {author} {\bibfnamefont {E.}~\bibnamefont {Lyman}},\ }\href@noop {} {\bibfield  {journal} {\bibinfo  {journal} {J. Am. Chem. Soc.}\ }\textbf {\bibinfo {volume} {136}},\ \bibinfo {pages} {725} (\bibinfo {year} {2014})}\BibitemShut {NoStop}%
\bibitem [{\citenamefont {Sodt}, \citenamefont {Pastor},\ and\ \citenamefont {Lyman}(2015)}]{sodt2015hexagonal}%
  \BibitemOpen
  \bibfield  {author} {\bibinfo {author} {\bibfnamefont {A.~J.}\ \bibnamefont {Sodt}}, \bibinfo {author} {\bibfnamefont {R.~W.}\ \bibnamefont {Pastor}}, \ and\ \bibinfo {author} {\bibfnamefont {E.}~\bibnamefont {Lyman}},\ }\href@noop {} {\bibfield  {journal} {\bibinfo  {journal} {Biophys. J.}\ }\textbf {\bibinfo {volume} {109}},\ \bibinfo {pages} {948} (\bibinfo {year} {2015})}\BibitemShut {NoStop}%
\bibitem [{\citenamefont {Cournia}, \citenamefont {Ullmann},\ and\ \citenamefont {Smith}(2007)}]{cournia2007differential}%
  \BibitemOpen
  \bibfield  {author} {\bibinfo {author} {\bibfnamefont {Z.}~\bibnamefont {Cournia}}, \bibinfo {author} {\bibfnamefont {G.~M.}\ \bibnamefont {Ullmann}}, \ and\ \bibinfo {author} {\bibfnamefont {J.~C.}\ \bibnamefont {Smith}},\ }\href@noop {} {\bibfield  {journal} {\bibinfo  {journal} {J. Phys. Chem. B}\ }\textbf {\bibinfo {volume} {111}},\ \bibinfo {pages} {1786} (\bibinfo {year} {2007})}\BibitemShut {NoStop}%
\bibitem [{\citenamefont {Hakobyan}\ and\ \citenamefont {Heuer}(2013)}]{hakobyan2013phase}%
  \BibitemOpen
  \bibfield  {author} {\bibinfo {author} {\bibfnamefont {D.}~\bibnamefont {Hakobyan}}\ and\ \bibinfo {author} {\bibfnamefont {A.}~\bibnamefont {Heuer}},\ }\href@noop {} {\bibfield  {journal} {\bibinfo  {journal} {J. Phys. Chem. B}\ }\textbf {\bibinfo {volume} {117}},\ \bibinfo {pages} {3841} (\bibinfo {year} {2013})}\BibitemShut {NoStop}%
\bibitem [{\citenamefont {Bennett}, \citenamefont {Shea},\ and\ \citenamefont {Tieleman}(2018)}]{bennett2018phospholipid}%
  \BibitemOpen
  \bibfield  {author} {\bibinfo {author} {\bibfnamefont {W.~D.}\ \bibnamefont {Bennett}}, \bibinfo {author} {\bibfnamefont {J.-E.}\ \bibnamefont {Shea}}, \ and\ \bibinfo {author} {\bibfnamefont {D.~P.}\ \bibnamefont {Tieleman}},\ }\href@noop {} {\bibfield  {journal} {\bibinfo  {journal} {Biophys. J.}\ }\textbf {\bibinfo {volume} {114}},\ \bibinfo {pages} {2595} (\bibinfo {year} {2018})}\BibitemShut {NoStop}%
\bibitem [{\citenamefont {Gu}, \citenamefont {Baoukina},\ and\ \citenamefont {Tieleman}(2020)}]{gu2020phase}%
  \BibitemOpen
  \bibfield  {author} {\bibinfo {author} {\bibfnamefont {R.-X.}\ \bibnamefont {Gu}}, \bibinfo {author} {\bibfnamefont {S.}~\bibnamefont {Baoukina}}, \ and\ \bibinfo {author} {\bibfnamefont {D.~P.}\ \bibnamefont {Tieleman}},\ }\href@noop {} {\bibfield  {journal} {\bibinfo  {journal} {J. Am. Chem. Soc.}\ }\textbf {\bibinfo {volume} {142}},\ \bibinfo {pages} {2844} (\bibinfo {year} {2020})}\BibitemShut {NoStop}%
\bibitem [{\citenamefont {Perlmutter}\ and\ \citenamefont {Sachs}(2009)}]{perlmutter2009inhibiting}%
  \BibitemOpen
  \bibfield  {author} {\bibinfo {author} {\bibfnamefont {J.~D.}\ \bibnamefont {Perlmutter}}\ and\ \bibinfo {author} {\bibfnamefont {J.~N.}\ \bibnamefont {Sachs}},\ }\href@noop {} {\bibfield  {journal} {\bibinfo  {journal} {J. Am. Chem. Soc.}\ }\textbf {\bibinfo {volume} {131}},\ \bibinfo {pages} {16362} (\bibinfo {year} {2009})}\BibitemShut {NoStop}%
\bibitem [{\citenamefont {Rosetti}\ and\ \citenamefont {Pastorino}(2012)}]{rosetti2012comparison}%
  \BibitemOpen
  \bibfield  {author} {\bibinfo {author} {\bibfnamefont {C.}~\bibnamefont {Rosetti}}\ and\ \bibinfo {author} {\bibfnamefont {C.}~\bibnamefont {Pastorino}},\ }\href@noop {} {\bibfield  {journal} {\bibinfo  {journal} {J. Phys. Chem. B}\ }\textbf {\bibinfo {volume} {116}},\ \bibinfo {pages} {3525} (\bibinfo {year} {2012})}\BibitemShut {NoStop}%
\bibitem [{\citenamefont {Davis}\ \emph {et~al.}(2013)\citenamefont {Davis}, \citenamefont {Sunil~Kumar}, \citenamefont {Sperotto},\ and\ \citenamefont {Laradji}}]{davis2013predictions}%
  \BibitemOpen
  \bibfield  {author} {\bibinfo {author} {\bibfnamefont {R.~S.}\ \bibnamefont {Davis}}, \bibinfo {author} {\bibfnamefont {P.}~\bibnamefont {Sunil~Kumar}}, \bibinfo {author} {\bibfnamefont {M.~M.}\ \bibnamefont {Sperotto}}, \ and\ \bibinfo {author} {\bibfnamefont {M.}~\bibnamefont {Laradji}},\ }\href@noop {} {\bibfield  {journal} {\bibinfo  {journal} {J. Phys. Chem. B}\ }\textbf {\bibinfo {volume} {117}},\ \bibinfo {pages} {4072} (\bibinfo {year} {2013})}\BibitemShut {NoStop}%
\bibitem [{\citenamefont {Baoukina}\ \emph {et~al.}(2013)\citenamefont {Baoukina}, \citenamefont {Mendez-Villuendas}, \citenamefont {Bennett},\ and\ \citenamefont {Tieleman}}]{baoukina2013computer}%
  \BibitemOpen
  \bibfield  {author} {\bibinfo {author} {\bibfnamefont {S.}~\bibnamefont {Baoukina}}, \bibinfo {author} {\bibfnamefont {E.}~\bibnamefont {Mendez-Villuendas}}, \bibinfo {author} {\bibfnamefont {W.~D.}\ \bibnamefont {Bennett}}, \ and\ \bibinfo {author} {\bibfnamefont {D.~P.}\ \bibnamefont {Tieleman}},\ }\href@noop {} {\bibfield  {journal} {\bibinfo  {journal} {Faraday Discuss.}\ }\textbf {\bibinfo {volume} {161}},\ \bibinfo {pages} {63} (\bibinfo {year} {2013})}\BibitemShut {NoStop}%
\bibitem [{\citenamefont {Pantelopulos}\ and\ \citenamefont {Straub}(2018)}]{pantelopulos2018regimes}%
  \BibitemOpen
  \bibfield  {author} {\bibinfo {author} {\bibfnamefont {G.~A.}\ \bibnamefont {Pantelopulos}}\ and\ \bibinfo {author} {\bibfnamefont {J.~E.}\ \bibnamefont {Straub}},\ }\href@noop {} {\bibfield  {journal} {\bibinfo  {journal} {Biophys. J.}\ }\textbf {\bibinfo {volume} {115}},\ \bibinfo {pages} {2167} (\bibinfo {year} {2018})}\BibitemShut {NoStop}%
\bibitem [{\citenamefont {Arnarez}\ \emph {et~al.}(2016)\citenamefont {Arnarez}, \citenamefont {Webb}, \citenamefont {Rouvi{\`e}re},\ and\ \citenamefont {Lyman}}]{arnarez2016hysteresis}%
  \BibitemOpen
  \bibfield  {author} {\bibinfo {author} {\bibfnamefont {C.}~\bibnamefont {Arnarez}}, \bibinfo {author} {\bibfnamefont {A.}~\bibnamefont {Webb}}, \bibinfo {author} {\bibfnamefont {E.}~\bibnamefont {Rouvi{\`e}re}}, \ and\ \bibinfo {author} {\bibfnamefont {E.}~\bibnamefont {Lyman}},\ }\href@noop {} {\bibfield  {journal} {\bibinfo  {journal} {J. Phys. Chem. B}\ }\textbf {\bibinfo {volume} {120}},\ \bibinfo {pages} {13086} (\bibinfo {year} {2016})}\BibitemShut {NoStop}%
\bibitem [{\citenamefont {Wang}\ \emph {et~al.}(2016)\citenamefont {Wang}, \citenamefont {Gkeka}, \citenamefont {Fuchs}, \citenamefont {Liedl},\ and\ \citenamefont {Cournia}}]{wang2016dppc}%
  \BibitemOpen
  \bibfield  {author} {\bibinfo {author} {\bibfnamefont {Y.}~\bibnamefont {Wang}}, \bibinfo {author} {\bibfnamefont {P.}~\bibnamefont {Gkeka}}, \bibinfo {author} {\bibfnamefont {J.~E.}\ \bibnamefont {Fuchs}}, \bibinfo {author} {\bibfnamefont {K.~R.}\ \bibnamefont {Liedl}}, \ and\ \bibinfo {author} {\bibfnamefont {Z.}~\bibnamefont {Cournia}},\ }\href@noop {} {\bibfield  {journal} {\bibinfo  {journal} {Biochim. Biophys. Acta, Biomembr.}\ }\textbf {\bibinfo {volume} {1858}},\ \bibinfo {pages} {2846} (\bibinfo {year} {2016})}\BibitemShut {NoStop}%
\bibitem [{\citenamefont {He}\ and\ \citenamefont {Maibaum}(2018)}]{he2018identifying}%
  \BibitemOpen
  \bibfield  {author} {\bibinfo {author} {\bibfnamefont {S.}~\bibnamefont {He}}\ and\ \bibinfo {author} {\bibfnamefont {L.}~\bibnamefont {Maibaum}},\ }\href@noop {} {\bibfield  {journal} {\bibinfo  {journal} {J. Phys. Chem. B}\ }\textbf {\bibinfo {volume} {122}},\ \bibinfo {pages} {3961} (\bibinfo {year} {2018})}\BibitemShut {NoStop}%
\bibitem [{\citenamefont {Carpenter}\ \emph {et~al.}(2018)\citenamefont {Carpenter}, \citenamefont {L{\'o}pez}, \citenamefont {Neale}, \citenamefont {Montour}, \citenamefont {Ing{\'o}lfsson}, \citenamefont {Di~Natale}, \citenamefont {Lightstone},\ and\ \citenamefont {Gnanakaran}}]{carpenter2018capturing}%
  \BibitemOpen
  \bibfield  {author} {\bibinfo {author} {\bibfnamefont {T.~S.}\ \bibnamefont {Carpenter}}, \bibinfo {author} {\bibfnamefont {C.~A.}\ \bibnamefont {L{\'o}pez}}, \bibinfo {author} {\bibfnamefont {C.}~\bibnamefont {Neale}}, \bibinfo {author} {\bibfnamefont {C.}~\bibnamefont {Montour}}, \bibinfo {author} {\bibfnamefont {H.~I.}\ \bibnamefont {Ing{\'o}lfsson}}, \bibinfo {author} {\bibfnamefont {F.}~\bibnamefont {Di~Natale}}, \bibinfo {author} {\bibfnamefont {F.~C.}\ \bibnamefont {Lightstone}}, \ and\ \bibinfo {author} {\bibfnamefont {S.}~\bibnamefont {Gnanakaran}},\ }\href@noop {} {\bibfield  {journal} {\bibinfo  {journal} {J. Chem. Theory Comput}\ }\textbf {\bibinfo {volume} {14}},\ \bibinfo {pages} {6050} (\bibinfo {year} {2018})}\BibitemShut {NoStop}%
\bibitem [{\citenamefont {Podewitz}\ \emph {et~al.}(2018)\citenamefont {Podewitz}, \citenamefont {Wang}, \citenamefont {Gkeka}, \citenamefont {von Grafenstein}, \citenamefont {Liedl},\ and\ \citenamefont {Cournia}}]{podewitz2018phase}%
  \BibitemOpen
  \bibfield  {author} {\bibinfo {author} {\bibfnamefont {M.}~\bibnamefont {Podewitz}}, \bibinfo {author} {\bibfnamefont {Y.}~\bibnamefont {Wang}}, \bibinfo {author} {\bibfnamefont {P.}~\bibnamefont {Gkeka}}, \bibinfo {author} {\bibfnamefont {S.}~\bibnamefont {von Grafenstein}}, \bibinfo {author} {\bibfnamefont {K.~R.}\ \bibnamefont {Liedl}}, \ and\ \bibinfo {author} {\bibfnamefont {Z.}~\bibnamefont {Cournia}},\ }\href@noop {} {\bibfield  {journal} {\bibinfo  {journal} {J. Phys. Chem. B}\ }\textbf {\bibinfo {volume} {122}},\ \bibinfo {pages} {10505} (\bibinfo {year} {2018})}\BibitemShut {NoStop}%
\bibitem [{\citenamefont {Sarkar}\ and\ \citenamefont {Farago}(2021)}]{PhysRevResearch.3.L042030}%
  \BibitemOpen
  \bibfield  {author} {\bibinfo {author} {\bibfnamefont {T.}~\bibnamefont {Sarkar}}\ and\ \bibinfo {author} {\bibfnamefont {O.}~\bibnamefont {Farago}},\ }\href {\doibase 10.1103/PhysRevResearch.3.L042030} {\bibfield  {journal} {\bibinfo  {journal} {Phys. Rev. Res.}\ }\textbf {\bibinfo {volume} {3}},\ \bibinfo {pages} {L042030} (\bibinfo {year} {2021})}\BibitemShut {NoStop}%
\bibitem [{\citenamefont {Sarkar}\ and\ \citenamefont {Farago}(2023{\natexlab{a}})}]{D2SM01025A}%
  \BibitemOpen
  \bibfield  {author} {\bibinfo {author} {\bibfnamefont {T.}~\bibnamefont {Sarkar}}\ and\ \bibinfo {author} {\bibfnamefont {O.}~\bibnamefont {Farago}},\ }\href {\doibase 10.1039/D2SM01025A} {\bibfield  {journal} {\bibinfo  {journal} {Soft Matter}\ }\textbf {\bibinfo {volume} {19}},\ \bibinfo {pages} {2417} (\bibinfo {year} {2023}{\natexlab{a}})}\BibitemShut {NoStop}%
\bibitem [{\citenamefont {Meinhardt}\ and\ \citenamefont {Schmid}(2019)}]{meinhardt2019structure}%
  \BibitemOpen
  \bibfield  {author} {\bibinfo {author} {\bibfnamefont {S.}~\bibnamefont {Meinhardt}}\ and\ \bibinfo {author} {\bibfnamefont {F.}~\bibnamefont {Schmid}},\ }\href@noop {} {\bibfield  {journal} {\bibinfo  {journal} {Soft Matter}\ }\textbf {\bibinfo {volume} {15}},\ \bibinfo {pages} {1942} (\bibinfo {year} {2019})}\BibitemShut {NoStop}%
\bibitem [{\citenamefont {Almeida}(2011)}]{almeida2011simple}%
  \BibitemOpen
  \bibfield  {author} {\bibinfo {author} {\bibfnamefont {P.~F.}\ \bibnamefont {Almeida}},\ }\href@noop {} {\bibfield  {journal} {\bibinfo  {journal} {Biophys. J.}\ }\textbf {\bibinfo {volume} {100}},\ \bibinfo {pages} {420} (\bibinfo {year} {2011})}\BibitemShut {NoStop}%
\bibitem [{\citenamefont {Lingwood}\ and\ \citenamefont {Simons}(2010)}]{lingwood2010lipid}%
  \BibitemOpen
  \bibfield  {author} {\bibinfo {author} {\bibfnamefont {D.}~\bibnamefont {Lingwood}}\ and\ \bibinfo {author} {\bibfnamefont {K.}~\bibnamefont {Simons}},\ }\href@noop {} {\bibfield  {journal} {\bibinfo  {journal} {science}\ }\textbf {\bibinfo {volume} {327}},\ \bibinfo {pages} {46} (\bibinfo {year} {2010})}\BibitemShut {NoStop}%
\bibitem [{\citenamefont {Stone}\ \emph {et~al.}(2017)\citenamefont {Stone}, \citenamefont {Shelby}, \citenamefont {N{\'u}{\~n}ez}, \citenamefont {Wisser},\ and\ \citenamefont {Veatch}}]{stone2017protein}%
  \BibitemOpen
  \bibfield  {author} {\bibinfo {author} {\bibfnamefont {M.~B.}\ \bibnamefont {Stone}}, \bibinfo {author} {\bibfnamefont {S.~A.}\ \bibnamefont {Shelby}}, \bibinfo {author} {\bibfnamefont {M.~F.}\ \bibnamefont {N{\'u}{\~n}ez}}, \bibinfo {author} {\bibfnamefont {K.}~\bibnamefont {Wisser}}, \ and\ \bibinfo {author} {\bibfnamefont {S.~L.}\ \bibnamefont {Veatch}},\ }\href@noop {} {\bibfield  {journal} {\bibinfo  {journal} {elife}\ }\textbf {\bibinfo {volume} {6}},\ \bibinfo {pages} {e19891} (\bibinfo {year} {2017})}\BibitemShut {NoStop}%
\bibitem [{\citenamefont {Danylchuk}\ \emph {et~al.}(2020)\citenamefont {Danylchuk}, \citenamefont {Sezgin}, \citenamefont {Chabert},\ and\ \citenamefont {Klymchenko}}]{danylchuk2020redesigning}%
  \BibitemOpen
  \bibfield  {author} {\bibinfo {author} {\bibfnamefont {D.~I.}\ \bibnamefont {Danylchuk}}, \bibinfo {author} {\bibfnamefont {E.}~\bibnamefont {Sezgin}}, \bibinfo {author} {\bibfnamefont {P.}~\bibnamefont {Chabert}}, \ and\ \bibinfo {author} {\bibfnamefont {A.~S.}\ \bibnamefont {Klymchenko}},\ }\href@noop {} {\bibfield  {journal} {\bibinfo  {journal} {Anal. Chem.}\ }\textbf {\bibinfo {volume} {92}},\ \bibinfo {pages} {14798} (\bibinfo {year} {2020})}\BibitemShut {NoStop}%
\bibitem [{\citenamefont {Beck-Garc{\'\i}a}\ \emph {et~al.}(2015)\citenamefont {Beck-Garc{\'\i}a}, \citenamefont {Beck-Garc{\'\i}a}, \citenamefont {Bohler}, \citenamefont {Zorzin}, \citenamefont {Sezgin}, \citenamefont {Levental}, \citenamefont {Alarc{\'o}n},\ and\ \citenamefont {Schamel}}]{beck2015nanoclusters}%
  \BibitemOpen
  \bibfield  {author} {\bibinfo {author} {\bibfnamefont {K.}~\bibnamefont {Beck-Garc{\'\i}a}}, \bibinfo {author} {\bibfnamefont {E.}~\bibnamefont {Beck-Garc{\'\i}a}}, \bibinfo {author} {\bibfnamefont {S.}~\bibnamefont {Bohler}}, \bibinfo {author} {\bibfnamefont {C.}~\bibnamefont {Zorzin}}, \bibinfo {author} {\bibfnamefont {E.}~\bibnamefont {Sezgin}}, \bibinfo {author} {\bibfnamefont {I.}~\bibnamefont {Levental}}, \bibinfo {author} {\bibfnamefont {B.}~\bibnamefont {Alarc{\'o}n}}, \ and\ \bibinfo {author} {\bibfnamefont {W.~W.}\ \bibnamefont {Schamel}},\ }\href@noop {} {\bibfield  {journal} {\bibinfo  {journal} {Biochim. Biophys. Acta, Mol. Cell Res.}\ }\textbf {\bibinfo {volume} {1853}},\ \bibinfo {pages} {802} (\bibinfo {year} {2015})}\BibitemShut {NoStop}%
\bibitem [{\citenamefont {Magnani}\ \emph {et~al.}(2004)\citenamefont {Magnani}, \citenamefont {Tate}, \citenamefont {Wynne}, \citenamefont {Williams},\ and\ \citenamefont {Haase}}]{magnani2004partitioning}%
  \BibitemOpen
  \bibfield  {author} {\bibinfo {author} {\bibfnamefont {F.}~\bibnamefont {Magnani}}, \bibinfo {author} {\bibfnamefont {C.~G.}\ \bibnamefont {Tate}}, \bibinfo {author} {\bibfnamefont {S.}~\bibnamefont {Wynne}}, \bibinfo {author} {\bibfnamefont {C.}~\bibnamefont {Williams}}, \ and\ \bibinfo {author} {\bibfnamefont {J.}~\bibnamefont {Haase}},\ }\href@noop {} {\bibfield  {journal} {\bibinfo  {journal} {J. Biol. Chem.}\ }\textbf {\bibinfo {volume} {279}},\ \bibinfo {pages} {38770} (\bibinfo {year} {2004})}\BibitemShut {NoStop}%
\bibitem [{\citenamefont {Cuddy}, \citenamefont {Winick-Ng},\ and\ \citenamefont {Rylett}(2014)}]{cuddy2014regulation}%
  \BibitemOpen
  \bibfield  {author} {\bibinfo {author} {\bibfnamefont {L.~K.}\ \bibnamefont {Cuddy}}, \bibinfo {author} {\bibfnamefont {W.}~\bibnamefont {Winick-Ng}}, \ and\ \bibinfo {author} {\bibfnamefont {R.~J.}\ \bibnamefont {Rylett}},\ }\href@noop {} {\bibfield  {journal} {\bibinfo  {journal} {J. Neurochem.}\ }\textbf {\bibinfo {volume} {128}},\ \bibinfo {pages} {725} (\bibinfo {year} {2014})}\BibitemShut {NoStop}%
\bibitem [{\citenamefont {Huang}\ \emph {et~al.}(1999)\citenamefont {Huang}, \citenamefont {Zhou}, \citenamefont {Feng}, \citenamefont {Lynch}, \citenamefont {Klumperman}, \citenamefont {DeArmond},\ and\ \citenamefont {Mobley}}]{huang1999nerve}%
  \BibitemOpen
  \bibfield  {author} {\bibinfo {author} {\bibfnamefont {C.-s.}\ \bibnamefont {Huang}}, \bibinfo {author} {\bibfnamefont {J.}~\bibnamefont {Zhou}}, \bibinfo {author} {\bibfnamefont {A.~K.}\ \bibnamefont {Feng}}, \bibinfo {author} {\bibfnamefont {C.~C.}\ \bibnamefont {Lynch}}, \bibinfo {author} {\bibfnamefont {J.}~\bibnamefont {Klumperman}}, \bibinfo {author} {\bibfnamefont {S.~J.}\ \bibnamefont {DeArmond}}, \ and\ \bibinfo {author} {\bibfnamefont {W.~C.}\ \bibnamefont {Mobley}},\ }\href@noop {} {\bibfield  {journal} {\bibinfo  {journal} {J. Biol. Chem.}\ }\textbf {\bibinfo {volume} {274}},\ \bibinfo {pages} {36707} (\bibinfo {year} {1999})}\BibitemShut {NoStop}%
\bibitem [{\citenamefont {Bodosa}, \citenamefont {Iyer},\ and\ \citenamefont {Srivastava}(2020{\natexlab{a}})}]{bodosa2020preferential}%
  \BibitemOpen
  \bibfield  {author} {\bibinfo {author} {\bibfnamefont {J.}~\bibnamefont {Bodosa}}, \bibinfo {author} {\bibfnamefont {S.~S.}\ \bibnamefont {Iyer}}, \ and\ \bibinfo {author} {\bibfnamefont {A.}~\bibnamefont {Srivastava}},\ }\href@noop {} {\bibfield  {journal} {\bibinfo  {journal} {J. Membr. Biol.}\ }\textbf {\bibinfo {volume} {253}},\ \bibinfo {pages} {551} (\bibinfo {year} {2020}{\natexlab{a}})}\BibitemShut {NoStop}%
\bibitem [{\citenamefont {Kedia}\ \emph {et~al.}(2020)\citenamefont {Kedia}, \citenamefont {Ramakrishna}, \citenamefont {Netrakanti}, \citenamefont {Jose}, \citenamefont {Sibarita}, \citenamefont {Nadkarni},\ and\ \citenamefont {Nair}}]{kedia2020real}%
  \BibitemOpen
  \bibfield  {author} {\bibinfo {author} {\bibfnamefont {S.}~\bibnamefont {Kedia}}, \bibinfo {author} {\bibfnamefont {P.}~\bibnamefont {Ramakrishna}}, \bibinfo {author} {\bibfnamefont {P.~R.}\ \bibnamefont {Netrakanti}}, \bibinfo {author} {\bibfnamefont {M.}~\bibnamefont {Jose}}, \bibinfo {author} {\bibfnamefont {J.-B.}\ \bibnamefont {Sibarita}}, \bibinfo {author} {\bibfnamefont {S.}~\bibnamefont {Nadkarni}}, \ and\ \bibinfo {author} {\bibfnamefont {D.}~\bibnamefont {Nair}},\ }\href@noop {} {\bibfield  {journal} {\bibinfo  {journal} {Nanoscale}\ }\textbf {\bibinfo {volume} {12}},\ \bibinfo {pages} {8200} (\bibinfo {year} {2020})}\BibitemShut {NoStop}%
\bibitem [{\citenamefont {Colin}\ \emph {et~al.}(2016)\citenamefont {Colin}, \citenamefont {Gregory-Pauron}, \citenamefont {Lanhers}, \citenamefont {Claudepierre}, \citenamefont {Corbier}, \citenamefont {Yen}, \citenamefont {Malaplate-Armand},\ and\ \citenamefont {Oster}}]{colin2016membrane}%
  \BibitemOpen
  \bibfield  {author} {\bibinfo {author} {\bibfnamefont {J.}~\bibnamefont {Colin}}, \bibinfo {author} {\bibfnamefont {L.}~\bibnamefont {Gregory-Pauron}}, \bibinfo {author} {\bibfnamefont {M.-C.}\ \bibnamefont {Lanhers}}, \bibinfo {author} {\bibfnamefont {T.}~\bibnamefont {Claudepierre}}, \bibinfo {author} {\bibfnamefont {C.}~\bibnamefont {Corbier}}, \bibinfo {author} {\bibfnamefont {F.~T.}\ \bibnamefont {Yen}}, \bibinfo {author} {\bibfnamefont {C.}~\bibnamefont {Malaplate-Armand}}, \ and\ \bibinfo {author} {\bibfnamefont {T.}~\bibnamefont {Oster}},\ }\href@noop {} {\bibfield  {journal} {\bibinfo  {journal} {Biochim.}\ }\textbf {\bibinfo {volume} {130}},\ \bibinfo {pages} {178} (\bibinfo {year} {2016})}\BibitemShut {NoStop}%
\bibitem [{\citenamefont {Sengupta}\ \emph {et~al.}(2019)\citenamefont {Sengupta}, \citenamefont {Seo}, \citenamefont {Pasolli}, \citenamefont {Song}, \citenamefont {Johnson},\ and\ \citenamefont {Lippincott-Schwartz}}]{sengupta2019lipid}%
  \BibitemOpen
  \bibfield  {author} {\bibinfo {author} {\bibfnamefont {P.}~\bibnamefont {Sengupta}}, \bibinfo {author} {\bibfnamefont {A.~Y.}\ \bibnamefont {Seo}}, \bibinfo {author} {\bibfnamefont {H.~A.}\ \bibnamefont {Pasolli}}, \bibinfo {author} {\bibfnamefont {Y.~E.}\ \bibnamefont {Song}}, \bibinfo {author} {\bibfnamefont {M.~C.}\ \bibnamefont {Johnson}}, \ and\ \bibinfo {author} {\bibfnamefont {J.}~\bibnamefont {Lippincott-Schwartz}},\ }\href@noop {} {\bibfield  {journal} {\bibinfo  {journal} {Nat. Cell Biol}\ }\textbf {\bibinfo {volume} {21}},\ \bibinfo {pages} {452} (\bibinfo {year} {2019})}\BibitemShut {NoStop}%
\bibitem [{\citenamefont {Yang}\ \emph {et~al.}(2017)\citenamefont {Yang}, \citenamefont {Kreutzberger}, \citenamefont {Kiessling}, \citenamefont {Ganser-Pornillos}, \citenamefont {White},\ and\ \citenamefont {Tamm}}]{yang2017hiv}%
  \BibitemOpen
  \bibfield  {author} {\bibinfo {author} {\bibfnamefont {S.-T.}\ \bibnamefont {Yang}}, \bibinfo {author} {\bibfnamefont {A.~J.}\ \bibnamefont {Kreutzberger}}, \bibinfo {author} {\bibfnamefont {V.}~\bibnamefont {Kiessling}}, \bibinfo {author} {\bibfnamefont {B.~K.}\ \bibnamefont {Ganser-Pornillos}}, \bibinfo {author} {\bibfnamefont {J.~M.}\ \bibnamefont {White}}, \ and\ \bibinfo {author} {\bibfnamefont {L.~K.}\ \bibnamefont {Tamm}},\ }\href@noop {} {\bibfield  {journal} {\bibinfo  {journal} {Sci. Adv.}\ }\textbf {\bibinfo {volume} {3}},\ \bibinfo {pages} {e1700338} (\bibinfo {year} {2017})}\BibitemShut {NoStop}%
\bibitem [{\citenamefont {Yang}, \citenamefont {Kiessling},\ and\ \citenamefont {Tamm}(2016)}]{yang2016line}%
  \BibitemOpen
  \bibfield  {author} {\bibinfo {author} {\bibfnamefont {S.-T.}\ \bibnamefont {Yang}}, \bibinfo {author} {\bibfnamefont {V.}~\bibnamefont {Kiessling}}, \ and\ \bibinfo {author} {\bibfnamefont {L.~K.}\ \bibnamefont {Tamm}},\ }\href@noop {} {\bibfield  {journal} {\bibinfo  {journal} {Nat. Commun.}\ }\textbf {\bibinfo {volume} {7}},\ \bibinfo {pages} {11401} (\bibinfo {year} {2016})}\BibitemShut {NoStop}%
\bibitem [{\citenamefont {Sch{\"a}fer}\ \emph {et~al.}(2011)\citenamefont {Sch{\"a}fer}, \citenamefont {de~Jong}, \citenamefont {Holt}, \citenamefont {Rzepiela}, \citenamefont {de~Vries}, \citenamefont {Poolman}, \citenamefont {Killian},\ and\ \citenamefont {Marrink}}]{schafer2011lipid}%
  \BibitemOpen
  \bibfield  {author} {\bibinfo {author} {\bibfnamefont {L.~V.}\ \bibnamefont {Sch{\"a}fer}}, \bibinfo {author} {\bibfnamefont {D.~H.}\ \bibnamefont {de~Jong}}, \bibinfo {author} {\bibfnamefont {A.}~\bibnamefont {Holt}}, \bibinfo {author} {\bibfnamefont {A.~J.}\ \bibnamefont {Rzepiela}}, \bibinfo {author} {\bibfnamefont {A.~H.}\ \bibnamefont {de~Vries}}, \bibinfo {author} {\bibfnamefont {B.}~\bibnamefont {Poolman}}, \bibinfo {author} {\bibfnamefont {J.~A.}\ \bibnamefont {Killian}}, \ and\ \bibinfo {author} {\bibfnamefont {S.~J.}\ \bibnamefont {Marrink}},\ }\href@noop {} {\bibfield  {journal} {\bibinfo  {journal} {Proc. Natl. Acad. Sci.}\ }\textbf {\bibinfo {volume} {108}},\ \bibinfo {pages} {1343} (\bibinfo {year} {2011})}\BibitemShut {NoStop}%
\bibitem [{\citenamefont {Kaiser}\ \emph {et~al.}(2011)\citenamefont {Kaiser}, \citenamefont {Or{\l}owski}, \citenamefont {R{\'o}g}, \citenamefont {Nyholm}, \citenamefont {Chai}, \citenamefont {Feizi}, \citenamefont {Lingwood}, \citenamefont {Vattulainen},\ and\ \citenamefont {Simons}}]{kaiser2011lateral}%
  \BibitemOpen
  \bibfield  {author} {\bibinfo {author} {\bibfnamefont {H.-J.}\ \bibnamefont {Kaiser}}, \bibinfo {author} {\bibfnamefont {A.}~\bibnamefont {Or{\l}owski}}, \bibinfo {author} {\bibfnamefont {T.}~\bibnamefont {R{\'o}g}}, \bibinfo {author} {\bibfnamefont {T.~K.}\ \bibnamefont {Nyholm}}, \bibinfo {author} {\bibfnamefont {W.}~\bibnamefont {Chai}}, \bibinfo {author} {\bibfnamefont {T.}~\bibnamefont {Feizi}}, \bibinfo {author} {\bibfnamefont {D.}~\bibnamefont {Lingwood}}, \bibinfo {author} {\bibfnamefont {I.}~\bibnamefont {Vattulainen}}, \ and\ \bibinfo {author} {\bibfnamefont {K.}~\bibnamefont {Simons}},\ }\href@noop {} {\bibfield  {journal} {\bibinfo  {journal} {Proc. Natl. Acad. Sci.}\ }\textbf {\bibinfo {volume} {108}},\ \bibinfo {pages} {16628} (\bibinfo {year} {2011})}\BibitemShut {NoStop}%
\bibitem [{\citenamefont {Periole}\ \emph {et~al.}(2007)\citenamefont {Periole}, \citenamefont {Huber}, \citenamefont {Marrink},\ and\ \citenamefont {Sakmar}}]{periole2007g}%
  \BibitemOpen
  \bibfield  {author} {\bibinfo {author} {\bibfnamefont {X.}~\bibnamefont {Periole}}, \bibinfo {author} {\bibfnamefont {T.}~\bibnamefont {Huber}}, \bibinfo {author} {\bibfnamefont {S.-J.}\ \bibnamefont {Marrink}}, \ and\ \bibinfo {author} {\bibfnamefont {T.~P.}\ \bibnamefont {Sakmar}},\ }\href@noop {} {\bibfield  {journal} {\bibinfo  {journal} {J. Am. Chem. Soc.}\ }\textbf {\bibinfo {volume} {129}},\ \bibinfo {pages} {10126} (\bibinfo {year} {2007})}\BibitemShut {NoStop}%
\bibitem [{\citenamefont {Domański}, \citenamefont {Marrink},\ and\ \citenamefont {Schäfer}(2012)}]{DOMANSKI2012984}%
  \BibitemOpen
  \bibfield  {author} {\bibinfo {author} {\bibfnamefont {J.}~\bibnamefont {Domański}}, \bibinfo {author} {\bibfnamefont {S.~J.}\ \bibnamefont {Marrink}}, \ and\ \bibinfo {author} {\bibfnamefont {L.~V.}\ \bibnamefont {Schäfer}},\ }\href {\doibase https://doi.org/10.1016/j.bbamem.2011.08.021} {\bibfield  {journal} {\bibinfo  {journal} {Biochim. Biophys. Acta, Biomembr.}\ }\textbf {\bibinfo {volume} {1818}},\ \bibinfo {pages} {984} (\bibinfo {year} {2012})},\ \bibinfo {note} {protein Folding in Membranes}\BibitemShut {NoStop}%
\bibitem [{\citenamefont {de~Jong}, \citenamefont {Lopez},\ and\ \citenamefont {Marrink}(2013)}]{de2013molecular}%
  \BibitemOpen
  \bibfield  {author} {\bibinfo {author} {\bibfnamefont {D.~H.}\ \bibnamefont {de~Jong}}, \bibinfo {author} {\bibfnamefont {C.~A.}\ \bibnamefont {Lopez}}, \ and\ \bibinfo {author} {\bibfnamefont {S.~J.}\ \bibnamefont {Marrink}},\ }\href@noop {} {\bibfield  {journal} {\bibinfo  {journal} {Faraday Discuss.}\ }\textbf {\bibinfo {volume} {161}},\ \bibinfo {pages} {347} (\bibinfo {year} {2013})}\BibitemShut {NoStop}%
\bibitem [{\citenamefont {Janosi}\ \emph {et~al.}(2012)\citenamefont {Janosi}, \citenamefont {Li}, \citenamefont {Hancock},\ and\ \citenamefont {Gorfe}}]{janosi2012organization}%
  \BibitemOpen
  \bibfield  {author} {\bibinfo {author} {\bibfnamefont {L.}~\bibnamefont {Janosi}}, \bibinfo {author} {\bibfnamefont {Z.}~\bibnamefont {Li}}, \bibinfo {author} {\bibfnamefont {J.~F.}\ \bibnamefont {Hancock}}, \ and\ \bibinfo {author} {\bibfnamefont {A.~A.}\ \bibnamefont {Gorfe}},\ }\href@noop {} {\bibfield  {journal} {\bibinfo  {journal} {Proc. Natl. Acad. Sci.}\ }\textbf {\bibinfo {volume} {109}},\ \bibinfo {pages} {8097} (\bibinfo {year} {2012})}\BibitemShut {NoStop}%
\bibitem [{\citenamefont {Lorent}\ \emph {et~al.}(2020)\citenamefont {Lorent}, \citenamefont {Levental}, \citenamefont {Ganesan}, \citenamefont {Rivera-Longsworth}, \citenamefont {Sezgin}, \citenamefont {Doktorova}, \citenamefont {Lyman},\ and\ \citenamefont {Levental}}]{lorent2020plasma}%
  \BibitemOpen
  \bibfield  {author} {\bibinfo {author} {\bibfnamefont {J.}~\bibnamefont {Lorent}}, \bibinfo {author} {\bibfnamefont {K.~R.}\ \bibnamefont {Levental}}, \bibinfo {author} {\bibfnamefont {L.}~\bibnamefont {Ganesan}}, \bibinfo {author} {\bibfnamefont {G.}~\bibnamefont {Rivera-Longsworth}}, \bibinfo {author} {\bibfnamefont {E.}~\bibnamefont {Sezgin}}, \bibinfo {author} {\bibfnamefont {M.}~\bibnamefont {Doktorova}}, \bibinfo {author} {\bibfnamefont {E.}~\bibnamefont {Lyman}}, \ and\ \bibinfo {author} {\bibfnamefont {I.}~\bibnamefont {Levental}},\ }\href@noop {} {\bibfield  {journal} {\bibinfo  {journal} {Nat. Chem. Biol}\ }\textbf {\bibinfo {volume} {16}},\ \bibinfo {pages} {644} (\bibinfo {year} {2020})}\BibitemShut {NoStop}%
\bibitem [{\citenamefont {Lin}, \citenamefont {Gorfe},\ and\ \citenamefont {Levental}(2018)}]{lin2018protein}%
  \BibitemOpen
  \bibfield  {author} {\bibinfo {author} {\bibfnamefont {X.}~\bibnamefont {Lin}}, \bibinfo {author} {\bibfnamefont {A.~A.}\ \bibnamefont {Gorfe}}, \ and\ \bibinfo {author} {\bibfnamefont {I.}~\bibnamefont {Levental}},\ }\href@noop {} {\bibfield  {journal} {\bibinfo  {journal} {Biophys. J.}\ }\textbf {\bibinfo {volume} {114}},\ \bibinfo {pages} {1936} (\bibinfo {year} {2018})}\BibitemShut {NoStop}%
\bibitem [{\citenamefont {Bodosa}, \citenamefont {Iyer},\ and\ \citenamefont {Srivastava}(2020{\natexlab{b}})}]{bodosa2020}%
  \BibitemOpen
  \bibfield  {author} {\bibinfo {author} {\bibfnamefont {J.}~\bibnamefont {Bodosa}}, \bibinfo {author} {\bibfnamefont {S.~S.}\ \bibnamefont {Iyer}}, \ and\ \bibinfo {author} {\bibfnamefont {A.}~\bibnamefont {Srivastava}},\ }\href@noop {} {\bibfield  {journal} {\bibinfo  {journal} {J. Membr. Biol.}\ }\textbf {\bibinfo {volume} {253}},\ \bibinfo {pages} {551} (\bibinfo {year} {2020}{\natexlab{b}})}\BibitemShut {NoStop}%
\bibitem [{\citenamefont {Lorent}\ \emph {et~al.}(2017)\citenamefont {Lorent}, \citenamefont {Diaz-Rohrer}, \citenamefont {Lin}, \citenamefont {Spring}, \citenamefont {Gorfe}, \citenamefont {Levental},\ and\ \citenamefont {Levental}}]{lorent2017structural}%
  \BibitemOpen
  \bibfield  {author} {\bibinfo {author} {\bibfnamefont {J.~H.}\ \bibnamefont {Lorent}}, \bibinfo {author} {\bibfnamefont {B.}~\bibnamefont {Diaz-Rohrer}}, \bibinfo {author} {\bibfnamefont {X.}~\bibnamefont {Lin}}, \bibinfo {author} {\bibfnamefont {K.}~\bibnamefont {Spring}}, \bibinfo {author} {\bibfnamefont {A.~A.}\ \bibnamefont {Gorfe}}, \bibinfo {author} {\bibfnamefont {K.~R.}\ \bibnamefont {Levental}}, \ and\ \bibinfo {author} {\bibfnamefont {I.}~\bibnamefont {Levental}},\ }\href@noop {} {\bibfield  {journal} {\bibinfo  {journal} {Nat. Commun.}\ }\textbf {\bibinfo {volume} {8}},\ \bibinfo {pages} {1219} (\bibinfo {year} {2017})}\BibitemShut {NoStop}%
\bibitem [{\citenamefont {Levental}, \citenamefont {Grzybek},\ and\ \citenamefont {Simons}(2010)}]{levental2010greasing}%
  \BibitemOpen
  \bibfield  {author} {\bibinfo {author} {\bibfnamefont {I.}~\bibnamefont {Levental}}, \bibinfo {author} {\bibfnamefont {M.}~\bibnamefont {Grzybek}}, \ and\ \bibinfo {author} {\bibfnamefont {K.}~\bibnamefont {Simons}},\ }\href@noop {} {\bibfield  {journal} {\bibinfo  {journal} {Biochemistry}\ }\textbf {\bibinfo {volume} {49}},\ \bibinfo {pages} {6305} (\bibinfo {year} {2010})}\BibitemShut {NoStop}%
\bibitem [{\citenamefont {Nilsson}(2007)}]{NILSSON20071}%
  \BibitemOpen
  \bibfield  {author} {\bibinfo {author} {\bibfnamefont {C.~L.}\ \bibnamefont {Nilsson}},\ }in\ \href {\doibase https://doi.org/10.1016/B978-044453077-6/50002-8} {\emph {\bibinfo {booktitle} {Lectins}}},\ \bibinfo {editor} {edited by\ \bibinfo {editor} {\bibfnamefont {C.~L.}\ \bibnamefont {Nilsson}}}\ (\bibinfo  {publisher} {Elsevier Science B.V.},\ \bibinfo {address} {Amsterdam},\ \bibinfo {year} {2007})\ pp.\ \bibinfo {pages} {1--13}\BibitemShut {NoStop}%
\bibitem [{\citenamefont {Cebecauer}\ \emph {et~al.}(2018)\citenamefont {Cebecauer}, \citenamefont {Amaro}, \citenamefont {Jurkiewicz}, \citenamefont {Sarmento}, \citenamefont {Sachl}, \citenamefont {Cwiklik},\ and\ \citenamefont {Hof}}]{cebecauer2018membrane}%
  \BibitemOpen
  \bibfield  {author} {\bibinfo {author} {\bibfnamefont {M.}~\bibnamefont {Cebecauer}}, \bibinfo {author} {\bibfnamefont {M.}~\bibnamefont {Amaro}}, \bibinfo {author} {\bibfnamefont {P.}~\bibnamefont {Jurkiewicz}}, \bibinfo {author} {\bibfnamefont {M.~J.}\ \bibnamefont {Sarmento}}, \bibinfo {author} {\bibfnamefont {R.}~\bibnamefont {Sachl}}, \bibinfo {author} {\bibfnamefont {L.}~\bibnamefont {Cwiklik}}, \ and\ \bibinfo {author} {\bibfnamefont {M.}~\bibnamefont {Hof}},\ }\href@noop {} {\bibfield  {journal} {\bibinfo  {journal} {Chem. Rev.}\ }\textbf {\bibinfo {volume} {118}},\ \bibinfo {pages} {11259} (\bibinfo {year} {2018})}\BibitemShut {NoStop}%
\bibitem [{\citenamefont {Marsh}(2008)}]{marsh2008protein}%
  \BibitemOpen
  \bibfield  {author} {\bibinfo {author} {\bibfnamefont {D.}~\bibnamefont {Marsh}},\ }\href@noop {} {\bibfield  {journal} {\bibinfo  {journal} {Biochim. Biophys. Acta, Biomembr.}\ }\textbf {\bibinfo {volume} {1778}},\ \bibinfo {pages} {1545} (\bibinfo {year} {2008})}\BibitemShut {NoStop}%
\bibitem [{\citenamefont {Niemela}\ \emph {et~al.}(2010)\citenamefont {Niemela}, \citenamefont {Miettinen}, \citenamefont {Monticelli}, \citenamefont {Hammaren}, \citenamefont {Bjelkmar}, \citenamefont {Murtola}, \citenamefont {Lindahl},\ and\ \citenamefont {Vattulainen}}]{niemela2010membrane}%
  \BibitemOpen
  \bibfield  {author} {\bibinfo {author} {\bibfnamefont {P.~S.}\ \bibnamefont {Niemela}}, \bibinfo {author} {\bibfnamefont {M.~S.}\ \bibnamefont {Miettinen}}, \bibinfo {author} {\bibfnamefont {L.}~\bibnamefont {Monticelli}}, \bibinfo {author} {\bibfnamefont {H.}~\bibnamefont {Hammaren}}, \bibinfo {author} {\bibfnamefont {P.}~\bibnamefont {Bjelkmar}}, \bibinfo {author} {\bibfnamefont {T.}~\bibnamefont {Murtola}}, \bibinfo {author} {\bibfnamefont {E.}~\bibnamefont {Lindahl}}, \ and\ \bibinfo {author} {\bibfnamefont {I.}~\bibnamefont {Vattulainen}},\ }\href@noop {} {\bibfield  {journal} {\bibinfo  {journal} {J. Am. Chem. Soc.}\ }\textbf {\bibinfo {volume} {132}},\ \bibinfo {pages} {7574} (\bibinfo {year} {2010})}\BibitemShut {NoStop}%
\bibitem [{\citenamefont {Mouritsen}\ and\ \citenamefont {Bloom}(1984)}]{mouritsen1984mattress}%
  \BibitemOpen
  \bibfield  {author} {\bibinfo {author} {\bibfnamefont {O.~G.}\ \bibnamefont {Mouritsen}}\ and\ \bibinfo {author} {\bibfnamefont {M.}~\bibnamefont {Bloom}},\ }\href@noop {} {\bibfield  {journal} {\bibinfo  {journal} {Biophys. J.}\ }\textbf {\bibinfo {volume} {46}},\ \bibinfo {pages} {141} (\bibinfo {year} {1984})}\BibitemShut {NoStop}%
\bibitem [{\citenamefont {Kusumi}\ \emph {et~al.}(2005)\citenamefont {Kusumi}, \citenamefont {Nakada}, \citenamefont {Ritchie}, \citenamefont {Murase}, \citenamefont {Suzuki}, \citenamefont {Murakoshi}, \citenamefont {Kasai}, \citenamefont {Kondo},\ and\ \citenamefont {Fujiwara}}]{kusumi2005paradigm}%
  \BibitemOpen
  \bibfield  {author} {\bibinfo {author} {\bibfnamefont {A.}~\bibnamefont {Kusumi}}, \bibinfo {author} {\bibfnamefont {C.}~\bibnamefont {Nakada}}, \bibinfo {author} {\bibfnamefont {K.}~\bibnamefont {Ritchie}}, \bibinfo {author} {\bibfnamefont {K.}~\bibnamefont {Murase}}, \bibinfo {author} {\bibfnamefont {K.}~\bibnamefont {Suzuki}}, \bibinfo {author} {\bibfnamefont {H.}~\bibnamefont {Murakoshi}}, \bibinfo {author} {\bibfnamefont {R.~S.}\ \bibnamefont {Kasai}}, \bibinfo {author} {\bibfnamefont {J.}~\bibnamefont {Kondo}}, \ and\ \bibinfo {author} {\bibfnamefont {T.}~\bibnamefont {Fujiwara}},\ }\href@noop {} {\bibfield  {journal} {\bibinfo  {journal} {Annu. Rev. Biophys. Biomol. Struct.}\ }\textbf {\bibinfo {volume} {34}},\ \bibinfo {pages} {351} (\bibinfo {year} {2005})}\BibitemShut {NoStop}%
\bibitem [{\citenamefont {Sperotto}, \citenamefont {Ipsen},\ and\ \citenamefont {Mouritsen}(1989)}]{sperotto1989theory}%
  \BibitemOpen
  \bibfield  {author} {\bibinfo {author} {\bibfnamefont {M.~M.}\ \bibnamefont {Sperotto}}, \bibinfo {author} {\bibfnamefont {J.~H.}\ \bibnamefont {Ipsen}}, \ and\ \bibinfo {author} {\bibfnamefont {O.~G.}\ \bibnamefont {Mouritsen}},\ }\href@noop {} {\bibfield  {journal} {\bibinfo  {journal} {Cell Biophys}\ }\textbf {\bibinfo {volume} {14}},\ \bibinfo {pages} {79} (\bibinfo {year} {1989})}\BibitemShut {NoStop}%
\bibitem [{\citenamefont {Hinderliter}\ \emph {et~al.}(2001)\citenamefont {Hinderliter}, \citenamefont {Almeida}, \citenamefont {Creutz},\ and\ \citenamefont {Biltonen}}]{hinderliter2001domain}%
  \BibitemOpen
  \bibfield  {author} {\bibinfo {author} {\bibfnamefont {A.}~\bibnamefont {Hinderliter}}, \bibinfo {author} {\bibfnamefont {P.~F.}\ \bibnamefont {Almeida}}, \bibinfo {author} {\bibfnamefont {C.~E.}\ \bibnamefont {Creutz}}, \ and\ \bibinfo {author} {\bibfnamefont {R.~L.}\ \bibnamefont {Biltonen}},\ }\href@noop {} {\bibfield  {journal} {\bibinfo  {journal} {Biochemistry}\ }\textbf {\bibinfo {volume} {40}},\ \bibinfo {pages} {4181} (\bibinfo {year} {2001})}\BibitemShut {NoStop}%
\bibitem [{\citenamefont {Hoferer}\ \emph {et~al.}(2019)\citenamefont {Hoferer}, \citenamefont {Bonfanti}, \citenamefont {Taloni}, \citenamefont {La~Porta},\ and\ \citenamefont {Zapperi}}]{hoferer2019protein}%
  \BibitemOpen
  \bibfield  {author} {\bibinfo {author} {\bibfnamefont {M.}~\bibnamefont {Hoferer}}, \bibinfo {author} {\bibfnamefont {S.}~\bibnamefont {Bonfanti}}, \bibinfo {author} {\bibfnamefont {A.}~\bibnamefont {Taloni}}, \bibinfo {author} {\bibfnamefont {C.~A.}\ \bibnamefont {La~Porta}}, \ and\ \bibinfo {author} {\bibfnamefont {S.}~\bibnamefont {Zapperi}},\ }\href@noop {} {\bibfield  {journal} {\bibinfo  {journal} {Phys. Rev. E.}\ }\textbf {\bibinfo {volume} {100}},\ \bibinfo {pages} {042410} (\bibinfo {year} {2019})}\BibitemShut {NoStop}%
\bibitem [{\citenamefont {Sarkar}\ and\ \citenamefont {Farago}(2023{\natexlab{b}})}]{sarkar2023}%
  \BibitemOpen
  \bibfield  {author} {\bibinfo {author} {\bibfnamefont {T.}~\bibnamefont {Sarkar}}\ and\ \bibinfo {author} {\bibfnamefont {O.}~\bibnamefont {Farago}},\ }\href@noop {} {\bibfield  {journal} {\bibinfo  {journal} {Eur. Phys. J. E}\ }\textbf {\bibinfo {volume} {46}},\ \bibinfo {pages} {99} (\bibinfo {year} {2023}{\natexlab{b}})}\BibitemShut {NoStop}%
\bibitem [{\citenamefont {Veatch}\ and\ \citenamefont {Keller}(2005)}]{veatch2005miscibility}%
  \BibitemOpen
  \bibfield  {author} {\bibinfo {author} {\bibfnamefont {S.~L.}\ \bibnamefont {Veatch}}\ and\ \bibinfo {author} {\bibfnamefont {S.~L.}\ \bibnamefont {Keller}},\ }\href@noop {} {\bibfield  {journal} {\bibinfo  {journal} {Phys. Rev. Lett.}\ }\textbf {\bibinfo {volume} {94}},\ \bibinfo {pages} {148101} (\bibinfo {year} {2005})}\BibitemShut {NoStop}%
\bibitem [{\citenamefont {Veatch}\ and\ \citenamefont {Keller}(2003)}]{veatch2003separation}%
  \BibitemOpen
  \bibfield  {author} {\bibinfo {author} {\bibfnamefont {S.~L.}\ \bibnamefont {Veatch}}\ and\ \bibinfo {author} {\bibfnamefont {S.~L.}\ \bibnamefont {Keller}},\ }\href@noop {} {\bibfield  {journal} {\bibinfo  {journal} {Biophys. J.}\ }\textbf {\bibinfo {volume} {85}},\ \bibinfo {pages} {3074} (\bibinfo {year} {2003})}\BibitemShut {NoStop}%
\bibitem [{\citenamefont {Hoshen}\ and\ \citenamefont {Kopelman}(1976)}]{hoshen1976percolation}%
  \BibitemOpen
  \bibfield  {author} {\bibinfo {author} {\bibfnamefont {J.}~\bibnamefont {Hoshen}}\ and\ \bibinfo {author} {\bibfnamefont {R.}~\bibnamefont {Kopelman}},\ }\href@noop {} {\bibfield  {journal} {\bibinfo  {journal} {Phys. Rev. B.}\ }\textbf {\bibinfo {volume} {14}},\ \bibinfo {pages} {3438} (\bibinfo {year} {1976})}\BibitemShut {NoStop}%
\bibitem [{\citenamefont {Kusumi}\ \emph {et~al.}(2020)\citenamefont {Kusumi}, \citenamefont {Fujiwara}, \citenamefont {Tsunoyama}, \citenamefont {Kasai}, \citenamefont {Liu}, \citenamefont {Hirosawa}, \citenamefont {Kinoshita}, \citenamefont {Matsumori}, \citenamefont {Komura}, \citenamefont {Ando} \emph {et~al.}}]{kusumi2020defining}%
  \BibitemOpen
  \bibfield  {author} {\bibinfo {author} {\bibfnamefont {A.}~\bibnamefont {Kusumi}}, \bibinfo {author} {\bibfnamefont {T.~K.}\ \bibnamefont {Fujiwara}}, \bibinfo {author} {\bibfnamefont {T.~A.}\ \bibnamefont {Tsunoyama}}, \bibinfo {author} {\bibfnamefont {R.~S.}\ \bibnamefont {Kasai}}, \bibinfo {author} {\bibfnamefont {A.-A.}\ \bibnamefont {Liu}}, \bibinfo {author} {\bibfnamefont {K.~M.}\ \bibnamefont {Hirosawa}}, \bibinfo {author} {\bibfnamefont {M.}~\bibnamefont {Kinoshita}}, \bibinfo {author} {\bibfnamefont {N.}~\bibnamefont {Matsumori}}, \bibinfo {author} {\bibfnamefont {N.}~\bibnamefont {Komura}}, \bibinfo {author} {\bibfnamefont {H.}~\bibnamefont {Ando}},  \emph {et~al.},\ }\href@noop {} {\bibfield  {journal} {\bibinfo  {journal} {Traffic}\ }\textbf {\bibinfo {volume} {21}},\ \bibinfo {pages} {106} (\bibinfo {year} {2020})}\BibitemShut {NoStop}%
\bibitem [{\citenamefont {Rosetti}, \citenamefont {Mangiarotti},\ and\ \citenamefont {Wilke}(2017)}]{rosetti2017sizes}%
  \BibitemOpen
  \bibfield  {author} {\bibinfo {author} {\bibfnamefont {C.~M.}\ \bibnamefont {Rosetti}}, \bibinfo {author} {\bibfnamefont {A.}~\bibnamefont {Mangiarotti}}, \ and\ \bibinfo {author} {\bibfnamefont {N.}~\bibnamefont {Wilke}},\ }\href@noop {} {\bibfield  {journal} {\bibinfo  {journal} {Biochim. Biophys. Acta, Biomembr.}\ }\textbf {\bibinfo {volume} {1859}},\ \bibinfo {pages} {789} (\bibinfo {year} {2017})}\BibitemShut {NoStop}%
\bibitem [{\citenamefont {Sezgin}\ \emph {et~al.}(2017)\citenamefont {Sezgin}, \citenamefont {Levental}, \citenamefont {Mayor},\ and\ \citenamefont {Eggeling}}]{sezgin2017mystery}%
  \BibitemOpen
  \bibfield  {author} {\bibinfo {author} {\bibfnamefont {E.}~\bibnamefont {Sezgin}}, \bibinfo {author} {\bibfnamefont {I.}~\bibnamefont {Levental}}, \bibinfo {author} {\bibfnamefont {S.}~\bibnamefont {Mayor}}, \ and\ \bibinfo {author} {\bibfnamefont {C.}~\bibnamefont {Eggeling}},\ }\href@noop {} {\bibfield  {journal} {\bibinfo  {journal} {Nat. Rev. Mol. Cell Biol.}\ }\textbf {\bibinfo {volume} {18}},\ \bibinfo {pages} {361} (\bibinfo {year} {2017})}\BibitemShut {NoStop}%
\bibitem [{\citenamefont {Komura}\ \emph {et~al.}(2016)\citenamefont {Komura}, \citenamefont {Suzuki}, \citenamefont {Ando}, \citenamefont {Konishi}, \citenamefont {Koikeda}, \citenamefont {Imamura}, \citenamefont {Chadda}, \citenamefont {Fujiwara}, \citenamefont {Tsuboi}, \citenamefont {Sheng} \emph {et~al.}}]{komura2016raft}%
  \BibitemOpen
  \bibfield  {author} {\bibinfo {author} {\bibfnamefont {N.}~\bibnamefont {Komura}}, \bibinfo {author} {\bibfnamefont {K.~G.}\ \bibnamefont {Suzuki}}, \bibinfo {author} {\bibfnamefont {H.}~\bibnamefont {Ando}}, \bibinfo {author} {\bibfnamefont {M.}~\bibnamefont {Konishi}}, \bibinfo {author} {\bibfnamefont {M.}~\bibnamefont {Koikeda}}, \bibinfo {author} {\bibfnamefont {A.}~\bibnamefont {Imamura}}, \bibinfo {author} {\bibfnamefont {R.}~\bibnamefont {Chadda}}, \bibinfo {author} {\bibfnamefont {T.~K.}\ \bibnamefont {Fujiwara}}, \bibinfo {author} {\bibfnamefont {H.}~\bibnamefont {Tsuboi}}, \bibinfo {author} {\bibfnamefont {R.}~\bibnamefont {Sheng}},  \emph {et~al.},\ }\href@noop {} {\bibfield  {journal} {\bibinfo  {journal} {Nat. Chem. Biol}\ }\textbf {\bibinfo {volume} {12}},\ \bibinfo {pages} {402} (\bibinfo {year} {2016})}\BibitemShut {NoStop}%
\end{thebibliography}%
\end{document}